\documentclass[amsmath,amssymb,aps,prl,onecolumn,notitlepage,superscriptaddress,longbibliography]{revtex4-2}

\usepackage{graphicx} 
\usepackage{amsmath}
\usepackage{amsbsy}
\usepackage{cancel}
\usepackage{enumerate}
\usepackage{mathtools}
\usepackage{amsfonts}
\usepackage{amssymb,dsfont,physics}
\usepackage{nicefrac}
\usepackage{color}
\usepackage{tikz}
\usetikzlibrary{positioning}
\usetikzlibrary{automata,positioning}
\usetikzlibrary{fit,shapes.geometric}
\usetikzlibrary{decorations.pathmorphing}
\usetikzlibrary{arrows,matrix}
\usetikzlibrary{calc}
\usepackage{hyperref}
\usepackage{comment}
\usepackage{circuitikz}

\newcommand{\proj}[1]{\ket{#1}\!\bra{#1}}

\newcommand{\N}{\mathcal{N}}
\newcommand{\E}{\mathcal{E}}
\newcommand{\F}{\mathcal{F}}
\newcommand{\W}{\mathcal{\W}}
\newcommand{\EE}{\tilde{\mathcal{E}}}
\newcommand{\FF}{\tilde{\mathcal{F}}}

\newcommand{\ts}{\textsuperscript}

\newcommand{\ee}{\tilde{E}}
\newcommand{\ff}{\tilde{F}}
\newcommand{\kk}{\tilde{K}}
\newcommand{\com}[2]{\left[#1, #2\right]}
\newcommand{\acom}[2]{\left\{#1, #2\right\}}

\newcommand{\D}{\mathcal{D}}
\newcommand{\Sw}{\mathcal{S}}
\newcommand{\Id}{\mathds{1}}
\newcommand{\pg}{P_g}

\newcommand{\id}{\mathbf{Id}}

\newcommand{\acm}{\mathfrak{a}}
\newcommand{\cm}{\mathfrak{c}}

\begin{document}

\title{Enhancing Quantum State Discrimination with Indefinite Causal Order}
\author{Spiros Kechrimparis}
\email{skechrimparis@gmail.com}
\affiliation{School of Computational Sciences, Korea Institute for Advanced Study, Seoul 02455, South Korea}

\author{James Moran}
\affiliation{Quantum Universe Center, Korea Institute for Advanced Study, Seoul 02455, South Korea}

\author{Athena Karsa}
\affiliation{School of Physics \& Astronomy, University College London, London WC1E 6BT, United Kingdom}
\affiliation{Korea Research Institute of Standards and Science, Daejeon 34113, South Korea}

\author{Changhyoup Lee}
\affiliation{Korea Research Institute of Standards and Science, Daejeon 34113, South Korea}

\author{Hyukjoon Kwon}
\affiliation{School of Computational Sciences, Korea Institute for Advanced Study, Seoul 02455, South Korea}

 \begin{abstract}
   The standard quantum state discrimination problem can be understood as a communication scenario involving a sender and a receiver following these three steps: (i) the sender encodes information in pre-agreed quantum states, (ii) sends them over a noiseless channel, and (iii) the receiver decodes the information by performing appropriate measurements on the received states. In a practical setting, however, the channel is not only noisy but often also unknown, thus altering the states and making optimal decoding generally not possible.
    In this work, we study this noisy discrimination scenario using a protocol based on indefinite causal order. To this end, we consider the quantum switch and define its higher-order generalisations, which we call \emph{superswitches}.
    We find that, for certain channels and ensembles, the guessing probability can be significantly improved compared to both single- and multi-copy state discrimination.
 \end{abstract}


\maketitle

\section{Introduction}
Discriminating quantum states underlies many of the practical applications of quantum information theory. These include quantum communication \cite{7044953, Chen_2021}, cryptography \cite{bennett_quantum_2014a}, data hiding \cite{terhal_hiding_2001,divincenzo_quantum_2002}, quantum secret sharing \cite{PhysRevA.59.1829}, as well as quantum-inspired machine learning \cite{giuntini_multiclass_2023}.

Quantum state discrimination, pioneered by Helstrom \cite{helstrom_quantum_1969} and Holevo \cite{holevo_probabilistic_2011},  may be understood as a communication scenario between two parties that agree on an ensemble of $n$ states, any of which may be selected with some finite probability. In general, owing to the non-orthogonality of quantum states, the states in the ensemble cannot be perfectly distinguished from one another. The goal of quantum state discrimination is then to optimise the measurement on the receiver's end to determine with maximum quantum mechanically-allowed probability which state was sent. In theoretical treatments of the problem, the quantum channel through which the states are being transmitted is taken to be noiseless, i.e.~it is the identity channel. For most practical purposes, however, some noise will be introduced during the transmission process. This situation was studied in Refs.\@ \cite{kechrimparis_preserving_2018,kechrimparis_channel_2019}, where the authors allowed for a possibly unknown channel between the two parties, introducing the problem of unknown state discrimination. It was shown that there exists a protocol such that an optimal measurement can be preserved for the optimal discrimination of the noisy states and that, moreover, this protocol sometimes enhances the guessing probability. In view of the fact that the protocol was based on channel twirling \cite{gross_evenly_2007}, an instance of a \emph{supermap} \cite{chiribella_quantum_2008,chiribella_transforming_2008}, it is natural to ask whether other supermaps can perform better at the task. 

Recently, a supermap known as the \emph{quantum switch} \cite{chiribella_quantum_2013} has attracted a lot of attention in the literature, owing to the fact that many tasks which are impossible classically or quantum mechanically with standard operations, can be successfully performed using it.  The quantum switch works by superposing the sequential action of two quantum channels by coupling them to an ancilla qubit upon which a measurement is performed. Since in the quantum switch protocol we cannot conclude which channel was applied first, the quantum switch is often referred to as a supermap with \emph{indefinite causal order}. Advantages for various tasks using indefinite causal order have been reported in Refs.\@ \cite{ebler_enhanced_2018,chiribella_indefinite_2021,salek_quantum_2018,procopio_communication_2019,procopio_sending_2020,sazim_classical_2021,chiribella_quantum_2021,mukhopadhyay_superposition_2020,das_quantum_2022}.
The first experiment that verified the presence of indefinite causal order was performed in Ref.\@ \cite{rubino_experimental_2017}. Since then, many other attempts to implement indefinite causal order in practice were performed. For a review of the current status we direct the reader to Ref.\@ \cite{rozema_experimental_2024b}.

In this work, we apply the quantum switch to the problem of state discrimination and find that in many cases a significant improvement in guessing probability is achieved. Moreover, often the problem of requiring a redesign of the optimal measurement is also circumvented allowing for optimal unknown state We also define higher-order quantum switches, which we refer to as \emph{superswitches}, and show that in certain cases these can further improve the guessing probability. 

The manuscript is structured as follows. We begin by reviewing known facts on quantum state discrimination and the quantum switch. We define our protocol and examine the guessing probability for various ensembles of states and channels, comparing it to standard quantum state discrimination bounds.
We define higher-order quantum switches, a natural extension of the standard quantum switch, and show that they can outperform the quantum switch in certain cases. Finally, we compare the performance of these superswitches for general Pauli channels, as well as define and study superswitches in higher state-space dimensions.

\section{Preliminaries}
\subsection{Motivation and problem statement}
The problem of quantum state discrimination can be formulated as a communication scenario between two parties, say Alice and Bob, who have agreed on an ensemble of states $\Omega=\left\{q_i,\rho_i\right\}_{i=1}^n$, a collection of states $\rho_i$ that appear with \emph{a priori} probabilities $q_i$. Alice selects a state $\rho_i$ according to the \emph{a priori} probabilities $q_i$ and sends it to Bob through a possibly noisy channel $\E$. Bob designs an appropriate measurement scheme, described by some \emph{positive operator valued measure} (POVM), with the goal of identifying the label of the state. In \emph{minimum-error} quantum state discrimination the figure of merit is the average probability of successful identification. Whenever $\E=\mathbf{Id}$, that is, the channel is the identity map, $\mathbf{Id}$, we recover the standard minimum-error discrimination problem. Specifically, for an ensemble $\Omega$, the minimum-error discrimination problem is to find a POVM $\Pi=\{\Pi_i\}_{i=1}^n$ that maximises the average probability of identifying the state correctly, that is,
\begin{align}
    \pg =\max_{\Pi} \sum_i q_i \tr(\Pi_i \rho_i ) \,. \label{eq: def guessing probability}
\end{align}

In the case of an ensemble of two states, $\{q_i, \rho_i\}_{i=1,2}$, the optimal guessing probability is given by the Helstrom bound \cite{helstrom_quantum_1969},
\begin{align}
	\pg = \frac{1}{2}+\frac{1}{2} \norm{q_1 \rho_1 -q_2 \rho_2}_1 \,. \label{eq: Helstrom bound}
 \end{align}

Even though closed form solutions exist only in a number of cases, necessary and sufficient conditions exist for the optimal measurement:
\begin{align}
\sum_i q_i \rho_i \Pi_i-q_j\rho_j \geq 0 \,\,\, \forall j \,.   \label{eq: iff optimal measurement}
\end{align}
Often a second condition is also mentioned,
\begin{align}
    \Pi_i (q_i \rho_i-q_i\rho_j) \Pi_j =0 \,\,\, \forall i,j\,,
\end{align}
which follows from the first. For a more detailed review on state discrimination, we refer the reader to Refs.\@ \cite{barnett_quantum_2009,bae_quantum_2015,bergou_discriminating_2004}, We note that the optimal measurement is not unique and that, moreover, some of the states may never be associated with any measurement outcomes, that is, they may never 	be identified by an optimal measurement (see Appendix A).

If a channel between the two parties is noisy, described by a quantum channel $\E$ acting between them, such that whenever Alice sends state $\rho_i$, the state $\E(\rho_i)$ is received by Bob. This defines a new ensemble, $\Omega^{(\E)}=\left\{q_i,\E(\rho_i)\right\}_{i=1}^n$, consisting of the noisy states to be discriminated. It is clear that the optimal measurement $\Pi$ of the original ensemble $\Omega$ is no longer optimal for the new ensemble $\Omega^{(\E)}$, in general. Since a quantum channel does not increase the guessing probability, it always follows that
\begin{align}
    \pg\geq \pg^{(\E)}\,,
\end{align}
where $\pg^{(\E)}$ is the guessing probability of the ensemble $\Omega^{(\E)}$ with the new optimal measurement $\Pi^{(\E)}$ performed.

Even though a quantum channel described mathematically as a \emph{completely positive} and \emph{trace preserving} (CPTP) map cannot enhance the guessing probability, a supermap can. This was noticed in Refs.\@ \cite{kechrimparis_preserving_2018,kechrimparis_channel_2019} and it was shown that \emph{channel twirling} implemented by the use of a \emph{unitary 2-design} \cite{gross_evenly_2007} can enhance the guessing probability for certain ensembles and channels. Note that since the effect of twirling is to produce a depolarisation channel, if the noise we start with is already depolarising, no improvement can ever be achieved by the twirling protocol. This is not the case, however, in the protocol we will propose in this work. Moreover, it was shown that for certain ensembles and channels it can also preserve the optimality of a quantum measurement for discrimination, saving the need for performing state or process tomography, both of which are costly. Such \emph{optimal measurement preserving channels} were completely characterised in dimension two and partially in dimension three and higher \cite{kechrimparis_optimal_2020}. Furthermore, experimental verification of indefinitely causally ordered processes has been demonstrated \cite{rubino_experimental_2017}, suggesting that not only do quantum supermaps have potential theoretical advantages over standard quantum operations, as we will show in this work, but they may also be of practical importance in the near future.

\subsection{The quantum switch}
We now introduce and review some basic results regarding the quantum switch. The quantum switch is a supermap that superposes the ordering between the sequence of actions of two channels $\E$ and $\F$. The resulting channel is defined as
\begin{align}
    S_\omega(\E,\F) = \sum_{i,j}K_{ij} (\rho\otimes \omega)K_{ij}^\dag \,,
\end{align}
with the Kraus operators
\begin{align}
    K_{ij} = E_i F_j \otimes \proj{0}_C+F_j E_i \otimes \proj{1}_C \,,
\end{align}
where $E_i$ and $F_j$ denote the Kraus operators of the channels $\E$ and $\F$ respectively, and $\omega$ denotes the ancilla qubit that controls the order of the channels.
At this point in the analysis we focus on Pauli channels given their ubiquitous nature in quantum information \cite{RevModPhys.87.307,PhysRevA.94.052325} and theoretical ease of use which allows for deeper analytical treatment. Furthermore, by techniques such as twirling \cite{PhysRevLett.76.722,PhysRevA.54.3824}, it is possible to convert arbitrary noise channels into Pauli channels, thus giving their analysis farther reaching consequences. Let $\E_{\vec{p}} (\rho) = \sum_i p_i \sigma_i \rho \sigma_i$ and $\F_{\vec{q}} (\rho) = \sum_i q_i \sigma_i \rho \sigma_i$, be two Pauli channels in dimension two, where $\vec{p}=(p_0,p_1,p_2,p_3)$ and  $\vec{q}=(q_0,q_1,q_2,q_3)$ denote probability vectors, i.e.~$\sum_i p_i = \sum_i q_i=1$ and $\sigma_i \in \{I,X,Y,Z\}$ the Pauli matrices. By choosing the control qubit to be $\omega=\ketbra{+}$, the action of the quantum switch is given by \cite{guo_experimental_2020}
\begin{align}
	\Sw _{\ketbra{+}} (\E_{\vec{p}},\F_{\vec{q}} )= r_+ C_+ (\rho) \otimes \ketbra{+} +r_- C_- (\rho) \otimes \ketbra{-},
\end{align}
where $r_+, r_-$ are probabilities defined as $r_-= r_{12}+r_{23}+r_{31}$, with $r_{ij}= p_i q_j +q_i p_j$, and $r_+ =  1-r_-$, while the channels $C_+, C_-$ are 
\begin{align}
	C_+(\rho) = \frac{(\sum_{i=0}^3 r_{ii}/2)\rho +\sum_{i=1}^3 r_{0i} \sigma_i \rho \sigma_i}{r_+} \,,
\end{align}
and
\begin{align}
	C_- (\rho) = \frac{r_{23} X \rho X +r_{31} Y \rho Y+ r_{12} Z \rho Z  }{r_-}\,.
\end{align}
 In the special case where the two channels are the same Pauli channel, 
 \begin{align}
	 \E_{\vec{p}}=\F_{\vec{q}}= p_0\rho+p_1 X\rho X +p_2 Y\rho Y+ p_3 Z\rho Z\,,  \label{eq:Pauli channel}
 \end{align}
 we find
\begin{align}
    S_\omega (\E_{\vec{p}},\E_{\vec{p}}) = q_+ C_+ (\rho) \otimes \omega_+ + q_- C_- (\rho) \otimes \omega_- \,,
\end{align}
and the expressions for the channels $C_+, C_-$ become
\begin{align}
    C_+(\rho) &= \frac{(p_0^2+p_1^2+p_2^2+p_3^2)\rho+2p_0(p_1 X\rho X+p_2 Y\rho Y +p_3 Z \rho Z)}{q_+} \,, \notag \\
    C_-(\rho) &= \frac{2 p_1 p_2 Z\rho Z+ 2p_2 p_3 X\rho X + 2 p_3 p_1 Y \rho Y}{q_-} \,, \label{eq: C_+ and C_- channels}
\end{align}
with
\begin{align}
    q_- = 2(p_1 p_2 + p_2 p_3 +p_3 p_1) \,,\, \, \, q_+ = 1-q_-\,, 
\end{align}
and $\omega_+ =\omega$ and $\omega_- = Z\omega Z$. If we make the choice $\omega = \proj{+}$ for an initial state of the ancilla system, we obtain $\omega_{\pm} = \proj{\pm}$. Performing a measurement in the $ \proj{\pm}$ basis, the two channels $C_+, C_-$ can thus be fully separated. We note that our examples in the following sections will be of this special type where input channels are the same.

The interpretation regarding the ancilla is that it is operated by a communication provider who is the only party that has access to it. The communication provider then performs a measurement on the ancilla and communicates the outcome to Bob. Thus, the ancilla cannot be used to encode information by any of the parties. Many of the advantages that follow from the quantum switch, can be explained as `consuming' the coherence of the ancilla to produce effects that are otherwise impossible with standard quantum operations \cite{chiribella_indefinite_2021}.

\section{Enhancing discrimination using the quantum switch}

\begin{figure}[!h]
	\centering
	\scalebox{0.7}{
		
		\tikzset{meter/.append style={draw, inner sep=8, rectangle, font=\vphantom{A}, minimum width=25, line width=.6,
				path picture={\draw[black] ([shift={(.1,.3)}]path picture bounding box.south west) to[bend left=50] ([shift={(-.1,.3)}]path picture bounding box.south east);\draw[black,-latex] ([shift={(0,.1)}]path picture bounding box.south) -- ([shift={(.3,-.1)}]path picture bounding box.north);}}}
		
		\begin{circuitikz} [node distance={25mm}, thick,square/.style={regular polygon,regular polygon sides=4,minimum size=1cm}, main/.style={draw, circle,minimum size=0.5cm}, squareDashed/.style={dashed, regular polygon,regular polygon sides=4,minimum size=1cm}]
			
			\node[] (1) {$\rho_j$};
			\node[] (2) [below right= 3cm and 1cm of 1] {$\ket{\omega}$};
			\node[meter] (3) [right=3cm of 2] {};

			\node[square,draw] (5) [above right=0.5cm and 3cm of 1] {$\E$};
			\node[square,draw] (6) [below right=0.5cm and 3cm of 1] {$\E$};
			\node[squareDashed, draw, fit={(5) (6)}] (4) {};
			
			\coordinate(9) at ($(2)!0.55!(3)$);
			\node[] (10) [right=0.9cm of 1] {};
			\node[] (11) [right=5.5cm of 1] {};
			
			\node[] (7) [above right=1cm and 8cm of 1] {$C_+(\rho_j)$};
			\node[] (8) [below right=1cm and 8cm of 1] {$C_-(\rho_j)$};
			
			\node[meter] (12) [right=2cm of 7] {};
			\node[meter] (13) [right=2cm of 8] {};
			\node[] (14) [right=1cm of 12] {$`k`$};
			\node[] (15) [right=1cm of 13] {$`k`$};
			\node[] (16) [above=0.1cm of 12] {$\Pi_+$};
			\node[] (17) [above=0.1cm of 13] {$\Pi_-$};
			
			\draw[] (9) to (4);
			\draw[->] (1) to (4);
			\draw[->] (2) to (3);
			\draw[->] (4) to [out=0,in=180,looseness=0.8] node[align=center,midway,above=0.2cm]{$'+'$} (7);
			\draw[->] (4) to [out=0,in=180,looseness=0.8] node[align=center,midway,below=0.2cm]{$'-'$} (8);
			
			\draw[green, -] (10) to [out=0,in=180,looseness=1.9] node[align=center,midway,above,left] {} (5);
			\draw[green, -] (5) to [out=0,in=180,looseness=1.9] node[align=center,midway,above,left] {} (6);
			\draw[green, -] (6) to [out=0,in=180,looseness=1.9] node[align=center,midway,above,left] {} (11);
			
			\draw[red, -] (10) to [out=0,in=180,looseness=1.9] node[align=center,midway,above,left] {} (6);
			\draw[red, -] (6) to [out=0,in=180,looseness=1.9] node[align=center,midway,above,left] {} (5);
			\draw[red, -] (5) to [out=0,in=180,looseness=1.9] node[align=center,midway,above,left] {} (11);
			
			\draw[->] (7) to [out=0,in=180,looseness=0] node[align=center,midway,above] {} (12) ;
			\draw[->] (8) to [out=0,in=180,looseness=0] node[align=center,midway,above] {} (13) ;
			\draw[->] (12) to [out=0,in=180,looseness=0] node[align=center,midway,above] {} (14) ;
			\draw[->] (13) to [out=0,in=180,looseness=0] node[align=center,midway,above] {} (15) ;
			
			
		\end{circuitikz} 
	}
	
	\caption{A visual representation of the protocol. Two parties want two communicate and have access to a noisy channel $\E$.  Instead of state $\rho_j$ being sent over the channel directly, the quantum switch is used first superposing two applications of the channel, followed by measurements in the ancilla qubit. The outcome of the measurement is communicated by the network provider to Bob, who then applies an appropriate discriminating measurement and a guess $k$ of the true label of the state $j$ is made.}
	\label{Fig: protocol}	
\end{figure}
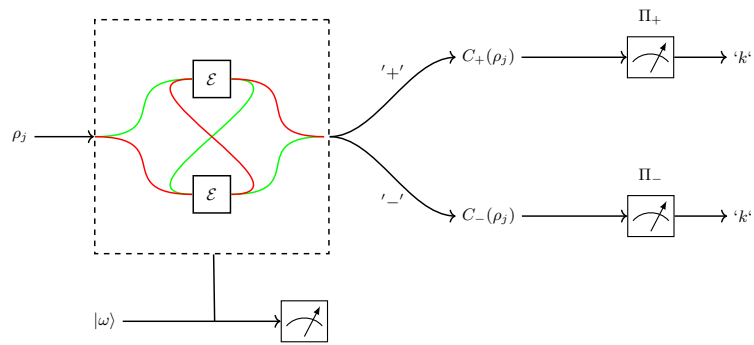
We now give some preliminary examples to demonstrate that an increase in guessing probability is possible in a discrimination scenario. The scenario works as follows. Alice prepares a copy of the state $\rho_j$ and sends it to the communication provider of the network. The communication provider feeds the state $\rho_j$ into the quantum switch that superposes two sequential applications of two channels $\E$ and $\F$ with $\F=\E$, and performs a measurement on the ancilla qubit. The ancilla measurement outcome is communicated to Bob who, depending on the measurement outcome on the ancilla, performs an appropriate measurement on $C_\pm(\rho_j)$ and subsequently makes a guess of the label of the state $\rho_j$, depending on the outcome of the measurement. A sketch of the protocol is shown in Fig.~\ref{Fig: protocol}. 

In principle, there are two scenarios in which the quantum switch can assist in increasing the guessing probability:
\begin{enumerate}
    \item Ensembles $\Omega$ and channels $\E$ such that channels $C_+, C_-$ preserve the optimal measurement or the new optimal measurement can be inferred from that of the original ensemble $\Omega$. As we will show, the depolarisation channel is a prototype of such a case. Then, the advantage of using the quantum switch is twofold: not only do we get an increase in guessing probability, but we also know what optimal measurement to apply without knowledge of the depolarisation parameter $p$. 
    
    \item Directly apply the optimal measurements for $C_+$ and $C_-$ (in general different from that of $\Omega$ or $\Omega^{(\E)}$) and then obtain the average guessing
    \begin{align}
        \pg^{(\Sw)}=q_+ \pg^+ + q_- \pg^-  \,. \label{eq: sep opt meas}
    \end{align}
    This is a case where we assume that we have performed process or state tomography to obtain information on the channel or states, and thus it is a scenario of enhancing communication on a given line with known noise. 
\end{enumerate}
We explore both instances in the following.

\subsection{Preliminary examples}
Here we demonstrate that increasing the guessing probability is theoretically possible using the quantum switch. Let $\N(\rho)$ be a Pauli channel of Eq.\@ \eqref{eq:Pauli channel} with $p_0=0$ and $p_1=p_2=p_3=\nicefrac{1}{3}$. Explicitly, $\N(\rho)=\D_{\nicefrac{4}{3}}(\rho)=\frac{1}{3}(X\rho X + Y\rho Y+  Z\rho Z)$, which is an instance of the depolarisation channel. We then readily obtain from Eq.\@ \eqref{eq: C_+ and C_- channels} that 
\begin{align}
    C_+(\rho) =\rho \, \,, \,\,\, C_-(\rho) = \N(\rho)\,,
    \label{eq:flipped POVM}
\end{align}
with $q_+=\nicefrac{1}{3}$ and $q_-=\nicefrac{2}{3}$.
Thus, the channel after a `+’ outcome is obtained upon a measurement of the ancilla is the identity, while the channel after a minus outcome is the noisy channel $\N$ itself. It follows that for any ensemble of states $\Omega$ with guessing probability $\pg$, Eqs.\@ \eqref{eq: def guessing probability} and \eqref{eq: sep opt meas} give
\begin{align}
    \pg^{(\Sw)}=\frac{1}{3}\pg+\frac{2}{3}\pg^{(\N)} \geq \pg^{(\N)} \,,
\end{align}
since $\pg \geq \pg ^{(\N)}$.

As a second example, consider a Pauli channel with $p_0=0$ and one of $p_1,p_2,p_3$ also equal to 0, $p_2=0$ say. Then,
$\E(\rho)=p X\rho X +(1-p)Z\rho Z$ and we find that $q_-=2p(1-p), q_+=1-2p(1-p)$, as well as $C_+(\rho)=\rho$ and $C_-(\rho)=Y\rho Y$, namely, one of the channels is the identity map and the other just a unitary. Thus, we can apply a $Y$ unitary in the case the `$-$' detector clicked at the control qubit to recover the original states, before performing the discrimination measurement. It follows that $\pg^{(\Sw)}=\pg$ for any ensemble. At the same time, the action of the channel $\E$ at the level of the Bloch vector is the following
\begin{align}
    \vec{r}=(r_1,r_2,r_3) \rightarrow \left(-(1-2p)r_1,-r_2,(1-2p)r_3\right)\,.
\end{align}
The channel flips the $y$ component and one of the $x$ or $z$ components depending on whether $p\in [0,\nicefrac{1}{2})$ or  $p\in (\nicefrac{1}{2},1]$, as well as shrinking the $x$ and $z$ components by a factor of $(1-2p)$. For $p=\nicefrac{1}{2}$ the Bloch sphere is squeezed onto the $y$ axis as well as mirrored over the origin. It follows that for any ensemble of states with Bloch vectors that do not lie only on the $y$ axis, the channel will have a decreased guessing probability compared to the one of the original ensemble, i.e.~$\pg^{(\E)}\leq \pg$, and thus the quantum switch always gives an advantage. Moreover, if all states of the ensemble lie on the $x$-$z$ plane and the value of the channel is $p=\nicefrac{1}{2}$, all states are sent to the maximally mixed state leading to complete loss of guessing probability. The protocol with the quantum switch can still recover the \emph{full} guessing probability, mirroring the effect in Ref.\@ \cite{chiribella_indefinite_2021} for the quantum capacity. This channel is unique up to unitaries.


\subsection{The depolarisation channel}
In this section we consider the case where the noise acting between the two parties is depolarising. Specifically we define the depolarisation channel as 
\begin{align}
    \D(\rho) = (1-p)\rho +p \frac{\Id}{2}\,, \,\, p\in[0,\nicefrac{4}{3}]\,, 
\end{align}
or equivalently, in the Kraus representation, as
\begin{align}
    \D(\rho) = \left(1-\frac{3p}{4}\right)\rho + \frac{p}{4}\left(X\rho X + Y\rho Y + Z\rho Z \right)\,, \,\, p\in[0,\nicefrac{4}{3}]\,. \label{eq:depolarisation}
\end{align}
The action of the depolarisation channel at the level of Bloch vectors is $\vec{r}\rightarrow (1-p)\vec{r}$. Specifically, as the parameter $p$ ranges in $[0,1)$, the vector is shrinking until the value $p=1$ where the map becomes completely depolarising and thus sends all states to the maximally mixed state. For values $p\in(1,\nicefrac{4}{3}]$ the Bloch vectors have flipped direction.
Consider a depolarisation channel given in Eq.\@ \eqref{eq:depolarisation}. Here, $p_0=1-\nicefrac{3p}{4},p_1=p_2=p_3=\nicefrac{p}{4}$, from which we obtain
\begin{align}
    q_-=\frac{3p^2}{8} \,, q_+=1-\frac{3p^2}{8}\,,
\end{align}
and
\begin{align}
    C_+(\rho)&=\left(1-\frac{3\tilde{p}}{4}\right)\rho + \frac{\tilde{p}}{4}\left(X\rho X + Y\rho Y + Z\rho Z \right)\,, \, \, \, \tilde{p}=\frac{4(4-3p)p}{8-3p^2}\,, \notag \\
    C_-(\rho)&=\frac{1}{3}\left(X\rho X + Y\rho Y + Z\rho Z \right) \,.
\end{align}
We see that both channels $C_\pm$ are themselves depolarisation channels. Note, that the original depolarisation channel has different optimal measurements depending on whether $p\in[0,1]$ or $p\in(1,\nicefrac{4}{3}]$ (see Appendix A). However, noting that $0<\tilde{p}\leq 1$, $C_+$ has the same optimal measurement $\Pi$ as the original ensemble, while $C_-$ has the measurement with flipped Bloch vectors $\tilde{\Pi}$, Eq.\@ \eqref{eq:flipped POVM}. Thus, even if we do not have information about the depolarisation parameter, $p$, of $\D_p$, the quantum switch allows us to always apply the optimal measurement and achieve the optimal guessing. This would not have been possible without state or process tomography otherwise, which highlights one of the benefits of employing the quantum switch for state discrimination. The second is the potential for increasing the guessing probability. Specifically, the guessing probabilities $\pg^+$ and $\pg^-$ for $C_+$ and $C_-$ respectively, are readily obtained from Eqs.\@ \eqref{eq:guessing prob depolarisation}-\eqref{eq:guessing prob depolarisation flipped} of Appendix A,
\begin{align}
    \pg^{(+)} = (1-\tilde{p})\pg+\frac{\tilde{p}}{n} \, \,, \,\,\,\,
    \pg^{(-)} = \frac{1}{3}\pg+\frac{2}{3n}\,,
\end{align}
where $n$ is the number of states in the ensemble. By setting the control qubit to $\omega=\proj{+}$ and performing a measurement on the $\ket{\pm}$ basis, we obtain the guessing probability for the quantum switch
\begin{align}
    \pg^{(\Sw)}=\left(1-\frac{3p^2}{8}\right)\left[\left(1-\frac{4(4-3p)p}{8-3p^2}\right)\pg+\frac{4(4-3p)p}{(8-3p^2)n}\right]+\frac{3p^2}{8}\left(\frac{1}{3}\pg+\frac{2}{3n}\right)\,. \label{eq:guessing switch depolarisation}
\end{align}
\begin{figure}[!t]
	\centering
	\includegraphics[width=0.7\linewidth]{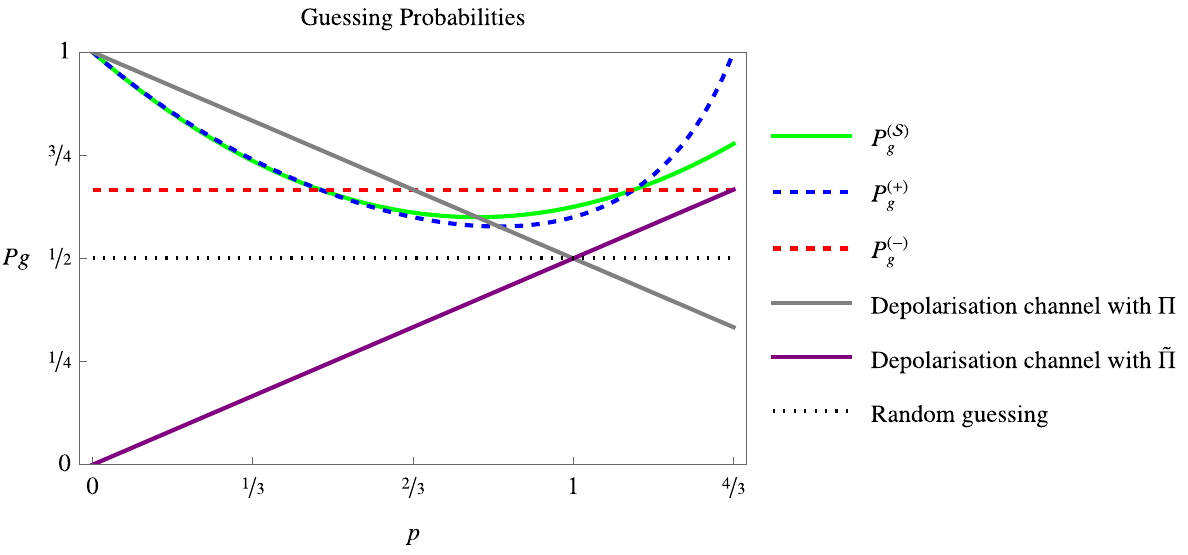}
	\caption{Guessing probabilities for the ensemble $\Omega$ consisting of an orthogonal pair of states with equal \emph{a priori} probabilities, sent through the depolarisation channel.  We show the graphs for the guessing probabilities for random guessing (dotted black), the quantum switch (green), a depolarisation channel with the original optimal measurement $\Pi$ (grey) and with the flipped measurement $\tilde{\Pi}$ (purple). We also plot $\pg^{(+)}$ (dashed blue) and $\pg^{(-)}$ (dashed red).}
	\label{fig:depolarisation}
\end{figure} 
In Fig.\@ \ref{fig:depolarisation}  we plot the guessing probability for the ensemble $\Omega=\{\nicefrac{1}{2},\ketbra{i}\}_{i=0,1}$ of two orthogonal states, i.e.~$\pg=1$, appearing with equal \emph{a priori} probabilities after sending them through the quantum switch and compare it to the guessing probability of the depolarisation channel. We see that for any value of $p>\nicefrac{4}{5}$, the quantum switch gives an advantage. Interestingly, at $p=1$ the depolarisation channel sends all states to the maximally mixed one, removing any possibility of guessing better than uniform, i.e.~$\pg=\nicefrac{1}{2}$, while the quantum switch allows for a correct detection with a probability of $\pg^{(\Sw)}=\nicefrac{5}{8}>\nicefrac{1}{2}$.

\subsection{Bit-phase flip channel}
Consider the bit-phase flip channel $\E(\rho)=(1-p)\rho+p Y\rho Y$, that is, a Pauli channel with $p_0=(1-p),p_2=p$ and $p_1=p_3=0$. Then, $q_- = 0, q_+ =1, C_-=0$ and
\begin{align}
    C_+ (\rho) = \left(p^2 + (1-p)^2\right)\rho +2p(1-p) Y\rho Y \,,
\end{align}
which is another bit-phase flip channel with $\hat{p}=2p(1-p)$. For values of $\hat{p}<p$, the portion of the identity is larger and thus the resulting contraction is smaller. This in principle will lead to better guessing but the optimal measurement cannot be directly inferred in general.
Since the action of the bit-phase flip channel with parameter $p$ is a uniform contraction along the $x-z$ plane of the Bloch ball by a factor of $1-2p$,
if the original ensemble of states all lie on the $x-y$ plane and appear with equal \emph{a priori} probabilities, then the channel acts as a depolarisation channel. If $p<\nicefrac{1}{2}$, the optimal measurement is preserved and we can readily compare the guessing probabilities. On the other hand, if $p>\nicefrac{1}{2}$, the Bloch vectors have flipped direction and so the measurement needs to be adjusted to $\tilde{\Pi}$, Eq.\@ \eqref{eq:flipped POVM}. The guessing probabilities are found as
\begin{align}
    \pg^{(\E)} &= (1-2p)\pg +\frac{2p}{n} \,, \notag \\
    \pg^{(\Sw)} &= (1-4p(1-p))\pg +\frac{4p(1-p)}{n} \,.
\end{align}
The first expression is optimal only for $p\in [0,\nicefrac{1}{2})$ while the second for all values of $p$ since $4(1-p)p>0$ and thus a flipping cannot occur.

\begin{figure}[!t]
	\centering	\includegraphics[width=0.6\linewidth]{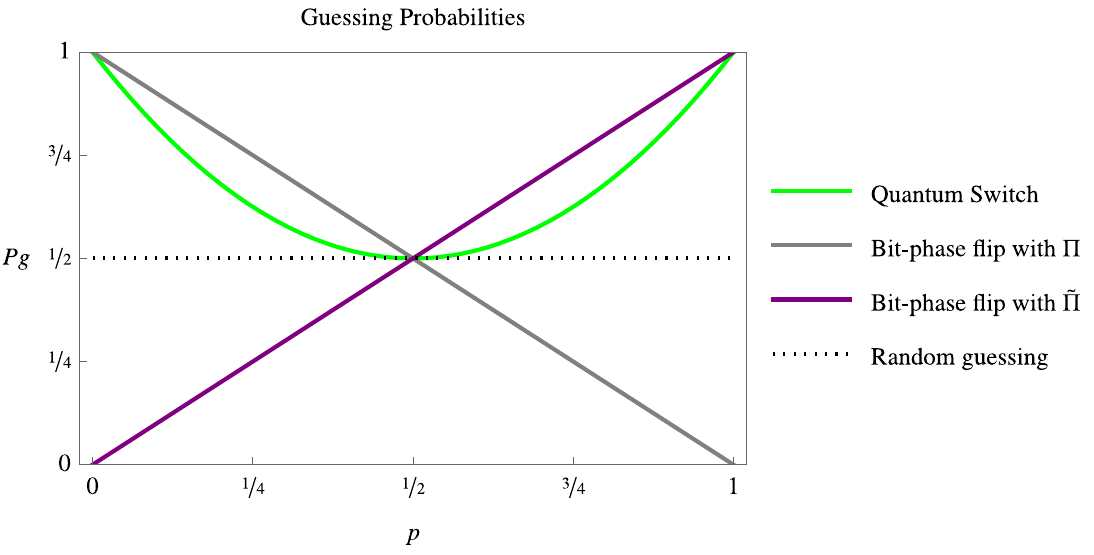}
	\caption{Guessing probabilities for an orthogonal pair of states lying on the $x-z$ plane and the bit-phase flip channel. We show the guessing probabilities for random guessing (dotted black), the quantum switch (green), the bit-phase flip channel with the original optimal measurement $\Pi$ (grey) and with the flipped measurement $\tilde{\Pi}$ (magenta).}
	\label{fig:bit-phase flip}
\end{figure} 

In Fig.\@ \ref{fig:bit-phase flip} we plot the guessing probabilities in all cases for any orthogonal pair of states lying on the $x-z$ plane. Note that if we do not have any information on the parameter range of $p$ in the bit-phase flip channel, then there is always a range where an improvement occurs by using the quantum switch. If, however, we assume that we know which measurement to apply after the action of $\E$, that is, we know whether $p\in[0,\nicefrac{1}{2})$ or  $p\in(\nicefrac{1}{2},1]$, then the quantum switch is always performing worse than the channel.

\subsection{Comparing to multiple-copy discrimination}
So far we have compared the quantum switch to single-copy state discrimination and observed that an advantage occurs in a number of cases. However, an objection could be raised since the quantum switch requires two applications of a channel $\E$ and thus it can be argued that this is the reason for the improvement. In this section, we compare the quantum switch to the state discrimination problem when two uses of the same channel are allowed. First, it is clear that two sequential uses of the same channel will only further degrade the guessing probability, unless the channel is unitary. Parallel use, however, will in principle offer a better guessing over a single copy scenario. Consider the two pure states $\rho_\pm = \alpha \ket{0} \pm \beta \ket{1}$, appearing with equal \emph{a priori} probabilities. Then,  $\pg = (1+\sqrt{1-c^2})/2$, where $c$ is the overlap between the states, $c=\abs{\alpha}^2-\abs{\beta}^2$.
If, however, we perform discrimination on $n$ copies we find the guessing $\pg^{(n)} = (1+\sqrt{1-c^{2n}})/2$, which is clearly larger than the case of single-copy discrimination. It should be noted that this comes at the cost of having to prepare two or more copies of the state, a cost that is not present in the case of the switch protocol.

\begin{figure}[!t]
	\centering	\includegraphics[width=0.6\linewidth]{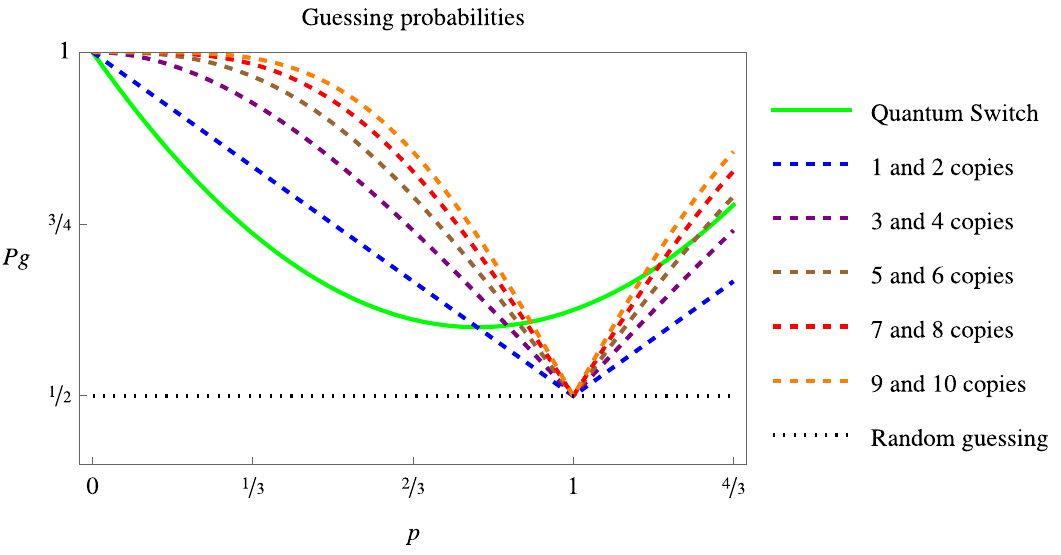}
	\caption{Guessing probabilities for an orthogonal pair of states for the quantum switch (green), a depolarisation channel for 1 and 2 copies (dashed blue), 3 and 4 copies (dashed magenta), 5 and 6 copies (dashed brown), 7 and 8 copies (dashed red), 9 and 10 copies (dashed orange) and random guessing (dotted black).}
	\label{fig:depolarisation 2 copy}
\end{figure}

We re-examine the case of the depolarisation channel and a pair of orthogonal states, $\rho_{i}=\ketbra{i}$ with $i=0,1$, appearing with equal \emph{a priori} probabilities. Given $n$ copies of the states after depolarisation noise  $\D_p(\rho_{i})$, with depolarisation strength $p$, this defines the ensemble $\Omega^{(n)}=\{\nicefrac{1}{2},\D_p(\rho_i)\otimes \underset{n}{\cdots} \otimes\D_p(\rho_i) \}_{i=0,1}$, for which the Helstrom bound becomes (see Appendix C)
 \begin{align}
     \pg^{(\D_p ,n)} &= \frac{1}{2}+\frac{1}{4}\norm{\D_p(\rho_0)\otimes \underset{n}{\cdots} \otimes\D_p(\rho_0)-\D_p(\rho_1)\otimes \underset{n}{\cdots} \otimes\D_p(\rho_1)}_1 \notag \\
     &=  \frac{1}{2}+\frac{1}{4} \sum_{k=0}^{n} \frac{n!}{k! (n-k)!}\abs{\left(\frac{p}{2}\right)^k \left(1-\frac{p}{2}\right)^{n-k}-\left(\frac{p}{2}\right)^{n-k} \left(1-\frac{p}{2}\right)^k}\,. \label{eq: Helstrom multiple orth depol}
 \end{align}
Note that the Helstrom bounds for $2n-1$ and $2n$ copies of states coincide. In Fig. \ref{fig:depolarisation 2 copy}, we plot the guessing probabilities for the discrimination of up to 10 copies. We see that there is always a region of improvement of the guessing probability with the quantum switch. However, this region becomes smaller with increasing number of copies. For any finite value $n$ there exists a region such that the protocol with the quantum switch outperforms the $n$-copy discrimination scenario. This is due to the fact that for near completely depolarising channels, that is depolarising channels with values of $p$ around the value $p=1$, the multiple-copy guessing probability is a continuous function of $p$ and has to approach random guessing. As $n$ tends to infinity, this region shrinks to the point $p=1$ where $\pg^{(\Sw)}=\nicefrac{5}{8}>\nicefrac{1}{2}$.

\section{First-order superswitch}

We now introduce higher-order switches that control the ordering of lower-order switches. Coherent superposition of the ordering of more than two channels was studied for certain informational quantities in Refs.\@ \cite{procopio_communication_2019,procopio_sending_2020,sazim_classical_2021,chiribella_quantum_2021,mukhopadhyay_superposition_2020} with the focus being on depolarisation channels. In Ref.\@ \cite{das_quantum_2022} the concept of switch of switches was introduced, similar to the case we consider here but in a restricted scenario where there is only a control switch for the inside switches, i.e.~their ordering is `synchronised'. In this work we will introduce the general case and we will show how it includes the previous approach as a special case at the end of the section. 

\begin{figure}[!h]
	\centering	\includegraphics[width=1\linewidth]{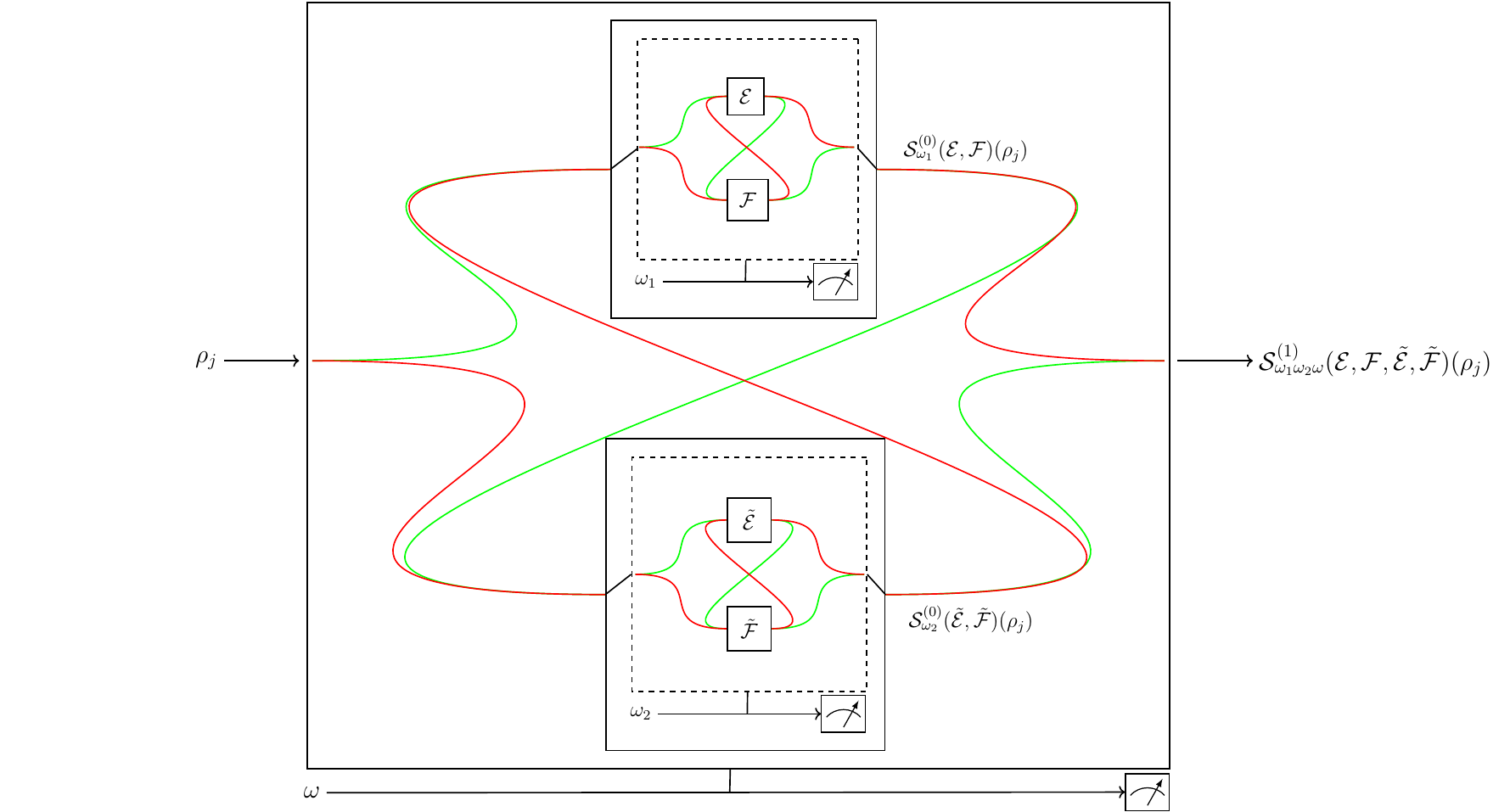}
	\caption{A visual representation of the first-order superswitch with control states being in a product state. Instead of state $\rho_j$ being sent into the quantum switch as in the protocol with one round, two switches are combined into the switch. Depending on the outcomes on measurements on the ancillas of the switches in the small boxes and the outcome of the switch in the large box, the outcome of the measurement is communicated by the network provider to Bob, who then applies an appropriate discriminating measurement and a guess $k$ of the true label of the state $j$ is made.}
	\label{Fig: first-order superswitch}	
\end{figure} 

It is instructive to discuss the first higher-order switch, which superimposes two quantum switches before discussing the general case. We call this the first-order superswitch and refer to the standard quantum switch as the zeroth-order superswitch. A schematic representation is shown in Fig. \ref{Fig: first-order superswitch}. Explicitly it is defined through 
\begin{align}
\Sw^{(1)}_{\omega_c}(\E,\F,\EE,\FF)(\rho)=\sum_{i,j,k,l} K_{ijkl}(\rho\otimes \omega_c)K_{ijkl}^\dagger \,,
\end{align}
where ancilla state $\omega$ control the order of channels $\E, \F$ and $\EE, \FF$ inside the two first-order switches respectively, as well as the ordering of the two switches themselves. The Kraus operators are explicitly given by
\begin{align}
	K_{ijkl}=&E_i F_j \ee_k \ff_l \otimes \ketbra{0}\otimes \ketbra{0}\otimes \ketbra{0} + E_i  F_j \ff_l \ee_k  \otimes \ketbra{0}\otimes \ketbra{1}\otimes \ketbra{0} \notag \\
	+& F_j E_i  \ee_k \ff_l \otimes \ketbra{1}\otimes \ketbra{0}\otimes \ketbra{0} +  F_j E_i \ff_l  \ee_k  \otimes \ketbra{1}\otimes \ketbra{1}\otimes \ketbra{0} \notag \\
	+&\ee_k \ff_l  E_i F_j \otimes \ketbra{0}\otimes \ketbra{0}\otimes \ketbra{1} + \ff_l \ee_k E_i  F_j  \otimes \ketbra{0}\otimes \ketbra{1}\otimes \ketbra{1} \notag \\
	+& \ee_k \ff_l  F_j E_i  \otimes \ketbra{1}\otimes \ketbra{0}\otimes \ketbra{1} + \ff_l  \ee_k  F_j E_i   \otimes \ketbra{1}\otimes \ketbra{1}\otimes \ketbra{1} \,, \label{eq: kraus 1st order superswitch}
\end{align} 
which can be rewritten as
\begin{align}
	K_{ijkl}=  \frac{1}{8} &\sum_{s_1, s_2, s =\pm} \Big[\big[E_i,F_j\big]_{s_1},\big[\ee_k,\ff_l\big]_{s_2}\Big]_s \otimes P_{s_1} \otimes P_{s_2} \otimes P_s=\notag \\
	\frac{1}{8}&\left( \acom{\acom{E_i}{F_j}}{\acom{\ee_k}{\ff_l}} \otimes \Id \otimes \Id \otimes \Id + \acom{\com{E_i}{F_j}}{\acom{\ee_k}{\ff_l}} \otimes Z\otimes \Id \otimes \Id \right.
	\notag \\
	&+\acom{\acom{E_i}{F_j}}{\com{\ee_k}{\ff_l}} \otimes \Id \otimes Z \otimes \Id +  \acom{\com{E_i}{F_j}}{\com{\ee_k}{\ff_l}} \otimes Z \otimes Z \otimes \Id \notag \\
	& + \com{\acom{E_i}{F_j}}{\acom{\ee_k}{\ff_l}} \otimes \Id \otimes \Id \otimes Z +\com{\com{E_i}{F_j}}{\acom{\ee_k}{\ff_l}}\otimes Z\otimes \Id \otimes Z
	\notag \\
	& \left.+\,\,\com{\acom{E_i}{F_j}}{\com{\ee_k}{\ff_l}} \otimes \Id \otimes Z \otimes Z + \,\, \com{\com{E_i}{F_j}}{\com{\ee_k}{\ff_l}}\otimes Z \otimes Z \otimes Z\right)\,,
\end{align}
where $[A,B]_{+} \equiv \{A,B\}=AB+BA$ and $[A,B]_{-} \equiv [A,B]=AB-BA$ denote the anticommutator and commutator, and $P_{+} =\Id$, $P_{-} = Z$. 
Under the assumption that for any pair of indices, both the commutator $[X_i,Y_j]$ and anticommutator $\{X_i,Y_j\}$ of any pair of Kraus operators $X_i, Y_j$  of the channels cannot be simultaneously non-zero (for instance Pauli channels), we arrive at the expression for the first-order superswitch,
\begin{align}
	\Sw^{(1)}_{\omega_c}(\E,\F,\EE,\FF)(\rho)=&\sum_{i,j,k,l} K_{ijkl}(\rho\otimes\omega_c)K_{ijkl}^\dagger= \frac{1}{64} \sum_{s_1, s_2, s =\pm} \Big[\big[E_i,F_j\big]_{s_1},\big[\ee_k,\ff_l\big]_{s_2}\Big]_s \rho \Big[\big[E_i,F_j\big]_{s_1},\big[\ee_k,\ff_l\big]_{s_2}\Big]_s^\dagger \notag \\
	 & \hspace{70mm}\otimes ( P_{s_1} \otimes P_{s_2} \otimes P_s) \omega_c ( P_{s_1} \otimes P_{s_2} \otimes P_s)^\dagger \,,
	 \label{eq: 1st order superswitch}
\end{align}
which when expanded gives
\begin{align}
	\Sw^{(1)}_{\omega_c}(\E,\F,\EE,\FF)(\rho)	=&\frac{1}{64}\sum_{i,j,k,l}\left(\,\, \acom{\acom{E_i}{F_j}}{\acom{\ee_k}{\ff_l}}\rho \acom{\acom{E_i}{F_j}}{\acom{\ee_k}{\ff_l}}^\dagger\otimes \omega_c  \right. \notag \\
	&\hspace{13mm}+\acom{\com{E_i}{F_j}}{\acom{\ee_k}{\ff_l}}\rho \acom{\com{E_i}{F_j}}{\acom{\ee_k}{\ff_l}}^\dagger\otimes (Z\otimes \Id \otimes\Id) \omega_c(Z\otimes \Id \otimes\Id)	\notag \\
	&\hspace{13mm}+\acom{\acom{E_i}{F_j}}{\com{\ee_k}{\ff_l}}\rho \acom{\acom{E_i}{F_j}}{\com{\ee_k}{\ff_l}}^\dagger\otimes (\Id\otimes Z \otimes\Id) \omega_c( \Id \otimes Z\otimes\Id)	\notag \\
	&\hspace{13mm}+\acom{\com{E_i}{F_j}}{\com{\ee_k}{\ff_l}}\rho \acom{\com{E_i}{F_j}}{\com{\ee_k}{\ff_l}}^\dagger\otimes (Z\otimes Z \otimes\Id) \omega_c (Z\otimes Z \otimes\Id)\notag \\
	&\hspace{13mm}+\com{\acom{E_i}{F_j}}{\acom{\ee_k}{\ff_l}}\rho \com{\acom{E_i}{F_j}}{\acom{\ee_k}{\ff_l}}^\dagger\otimes (\Id\otimes \Id \otimes Z) \omega_c (\Id \otimes \Id \otimes Z) \notag \\
	&\hspace{13mm}+\com{\com{E_i}{F_j}}{\acom{\ee_k}{\ff_l}}\rho \com{\com{E_i}{F_j}}{\acom{\ee_k}{\ff_l}}^\dagger(Z\otimes \Id \otimes Z) \omega_c (Z \otimes \Id \otimes Z)	\notag \\
	&\hspace{13mm}+\com{\acom{E_i}{F_j}}{\com{\ee_k}{\ff_l}}\rho \com{\acom{E_i}{F_j}}{\com{\ee_k}{\ff_l}}^\dagger(\Id\otimes Z \otimes Z) \omega_c (\Id \otimes Z \otimes Z)	\notag \\
	&\left.\hspace{13mm}+\com{\com{E_i}{F_j}}{\com{\ee_k}{\ff_l}}\rho \com{\com{E_i}{F_j}}{\com{\ee_k}{\ff_l}}^\dagger(Z\otimes Z \otimes Z) \omega_c (Z \otimes Z \otimes Z)	\right) \,. \label{eq: 1st order superswitch expanded}
\end{align}
We now assume that the ancilla state is a product state, i.e.~$\omega_c = \omega_1 \otimes \omega_2 \otimes \omega$, where $\omega_1$ and $\omega_2$ control the ordering of the channels $\E, \F$ and $\EE, \FF$, respectively, in the innermost switches and $\omega$ controls the ordering of the innermost switches themselves. Letting  $\omega_1=\omega_2=\omega=\ketbra{+}$ and defining channels $C_{ijk}^{(1)}$ as, e.g.
\begin{align}
	C^{(1)}_{+-+}(\E,\F,\EE,\FF)(\rho) = \frac{1}{r_{+-+}} \frac{1}{64}\sum_{i,j,k,l}\acom{\acom{E_i}{F_j}}{\com{\ee_k}{\ff_l}}\rho \acom{\acom{E_i}{F_j}}{\com{\ee_k}{\ff_l}}^\dagger\,,
\end{align}
where
\begin{align}
r_{+-+} =	\tr\left(\frac{1}{64 }\sum_{i,j,k,l}\acom{\acom{E_i}{F_j}}{\com{\ee_k}{\ff_l}}\rho \acom{\acom{E_i}{F_j}}{\com{\ee_k}{\ff_l}}^\dagger\right),
\end{align}
are the associated probabilities, we finally obtain
\begin{align}
	\Sw^{(1)}(\E,\F,\EE,\FF)&(\rho)= \sum_{s_1,s_2,s = \pm} r_{s_1 s_2 s } C^{(1)}_{s_1 s_2 s} (\E, \F, \EE, \FF) \otimes \ketbra{s_1 s_2 s}\notag \\
	= \,\,\,\,\,& r_{+++} C^{(1)}_{+++}(\E,\F,\EE,\FF)(\rho) \otimes\ketbra{+++}
	+r_{-++} C^{(1)}_{-++}(\E,\F,\EE,\FF)(\rho)\otimes \ketbra{-++}	\notag \\
	+&r_{+-+} C^{(1)}_{+-+}(\E,\F,\EE,\FF)(\rho)\otimes \ketbra{+-+}	+r_{--+} C^{(1)}_{--+}(\E,\F,\EE,\FF)(\rho)\otimes \ketbra{--+}	\notag \\
	+&r_{++-} C^{(1)}_{++-}(\E,\F,\EE,\FF)(\rho)\otimes\ketbra{++-}+r_{-+-} C^{(1)}_{-+-}(\E,\F,\EE,\FF)(\rho)\otimes \ketbra{-+-}	\notag \\
	+&r_{+--} C^{(1)}_{+--}(\E,\F,\EE,\FF)(\rho)\otimes\ketbra{+--}	+ r_{---} C^{(1)}_{---}(\E,\F,\EE,\FF)(\rho) \otimes \ketbra{---} \,.
\end{align}
It is clear that measurements on the three ancilla qubits allow for complete separation of the eight channels $C^{(1)}_{s_1 s_2 s}$, as in the case of the quantum switch. We note that if instead of taking a product state $\omega_1\otimes \omega _2 =\ketbra{+} \otimes \ketbra{+}$ for the two control qubits of the inside switches, we take one of the four entagled states, e.g. $\omega_{12}= \ketbra{\Phi^+}$ with $\ket{\Phi^+}=\frac{1}{\sqrt{2}}(\ket{00}+\ket{11})$, then our result reduces to the special case in Ref.\@ \cite{das_quantum_2022} (see Appendix D). The result of this is that the eight outcomes of the first-order superswitch are reduced to four, each one of them being a mixture of two of the outcomes of the general case introduced here. However, the mixing of two channels can never lead to higher guessing than mixing the guessing of the two individual channels: at most it can match it which happens in the case where the two channels share the same optimal measurement. In the case of ensembles of two states this follows from the triangle inequality of norms as the guessing probability is given from the Helstrom bound \cite{helstrom_quantum_1969}. For two channels $\E$ and $\F$ we have
\begin{align}
	\norm{\lambda \E(\rho) + (1-\lambda) \F(\rho)}_1 \leq \lambda \norm{\E(\rho)}_1 + (1-\lambda) \norm{\F(\rho)}_1 \,.
\end{align}
Thus, in general we have that 
\begin{align}
	\pg^{(\lambda \E + (1-\lambda) \F)}\leq \lambda \pg^{(\E)}+(1-\lambda)\pg^{(\F)} \,,
\end{align}
and it is clear that there is a trade-off between complexity of the setup and performance in a communication task such as state discrimination. 

By further assuming that all channels $\E,\F,\EE,\FF$ are equal to the same Pauli channel, $\E=\F=\EE=\FF= \sum_i p_i \sigma_i \rho \sigma_i$, we use the simplified notation $C^{(1)}_{ijk}(\rho)\equiv C^{(1)}_{ijk}(\E,\E,\E,\E)(\rho)$ since in this case there is no ambiguity about the channels. In the next section we examine a number of examples where the first-order superswitch can lead to better guessing probabilities.

\subsection{State discrimination with the first order superswitch}
 \begin{figure}[!t]
	\centering	\includegraphics[width=0.65\linewidth]{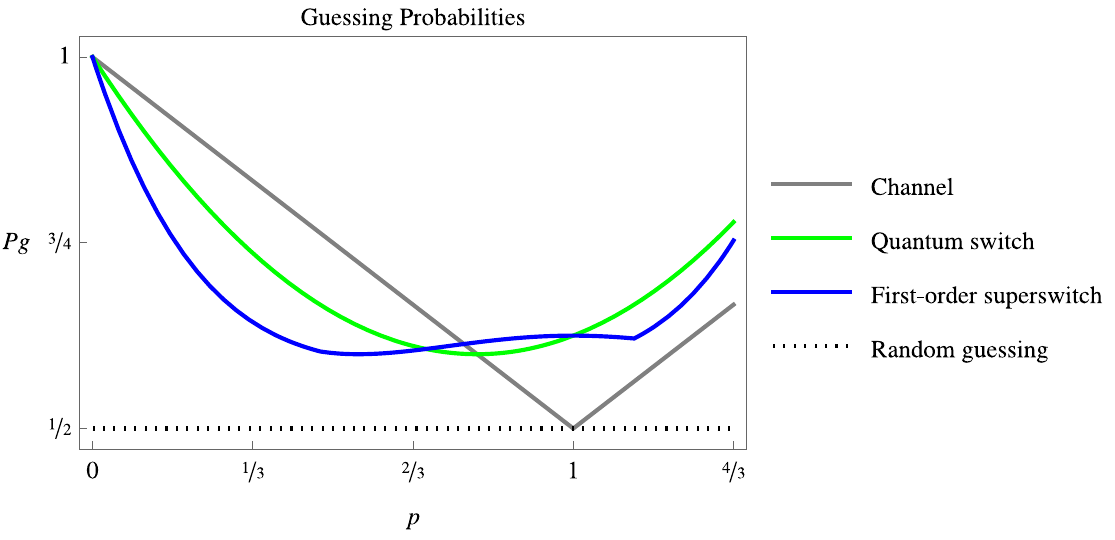}
	\caption{Guessing probabilities for an orthogonal pair of states for a depolarisation channel (grey), the quantum switch (green) and the first-order superswitch (blue).}
	\label{fig: depolarisation 1st}
\end{figure} 

We first re-examine the case of the depolarisation channel, that is $\E(\rho) = \D_p (\rho)$. We find that all eight possible channels of the first-order superswitch after measurements on the ancillas are instances of the depolarisation channel themselves. Explicitly, we obtain
\begin{align}
	&C^{(1)}_{+++}(\rho)=\D_{\eta_1} (\rho) \,, \,\, \mbox{with} \,\, \eta_1=1-\frac{99p^4-336p^3+432p^2-256p+64}{-45p^4+144p^3-144p^2+64} \,, \notag \\
	&C^{(1)}_{-++}(\rho)=C^{(1)}_{+-+}(\rho)=\D_{\eta_2} (\rho) \,, \,\, \mbox{with} \,\, \eta_2=1-\frac{-15p^2+24p-8}{9p^2-24p+24}\,, \notag \\
	&C^{(1)}_{--+}(\rho)=\rho \,, \,\, \mbox{and}	\,\,	C^{(1)}_{++-}(\rho)=C^{(1)}_{-+-}(\rho)=C^{(1)}_{+--}(\rho)=C^{(1)}_{---}(\rho)=\D_{\nicefrac{4}{3}} (\rho)\,.
\end{align}
The respective probabilities of occurrence are also found to be
\begin{align}
	r_{+++}&=1-\frac{9}{64}(5p^4-16p^3+16p^2)\,, \,\, r_{-++}=r_{+-+}=\frac{3}{64}\left(3p^4-8p^3+8p^2\right), \,\, r_{--+}=\frac{3p^2}{64} \,, \notag \\
	r_{++-}&=\frac{3}{32}(4-3p)^2 p^2\,, \,\, r_{-+-}=r_{+--}=\frac{3}{32}(4-3p) p^3\,, \,\, r_{---}=\frac{3}{32} p^4\,.
\end{align}

In Fig.\@ \ref{fig: depolarisation 1st} we plot the guessing probabilities for a depolarisation channel for the first-order superswitch and contrast it with the quantum switch. We find that there is intricate behaviour, with regions where both are performing worse than the channel, others where the first-order superswitch is outperforming the quantum switch and vice versa.

\begin{figure}[!b]
	\centering	\includegraphics[width=0.55\linewidth]{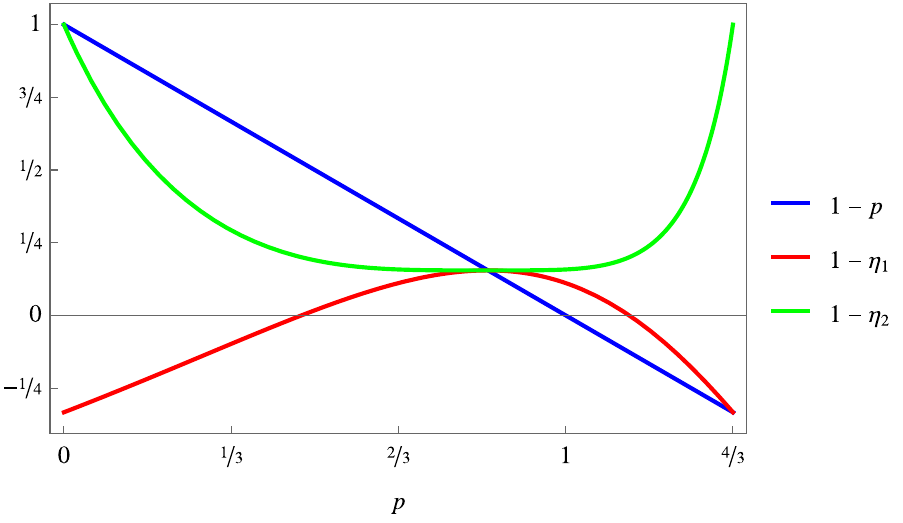}
	\caption{Plot of the parameters $1-p, 1-\eta_1,1-\eta_2$ that control the direction of the Bloch vectors of the input states of the channels. Negative values imply that the Bloch vectors have flipped direction.}
	\label{fig: depolarisation no OMP}
\end{figure}

We note that even though there is a region where the first-order superswitch outperforms the quantum switch, this comes at a cost. Remembering that one of the advantages of the quantum switch in the case of the depolarisation channel was that one could always infer the optimal measurement to be applied depending on the outcome of the measurement on the ancilla, this does not hold true in the case of the higher-order superswitches.  In Fig.\@ \ref{fig: depolarisation no OMP} we plot the parameters $1-\eta_1,1-\eta_2$ that control the direction of the Bloch vectors of the input states of the channels, depending on whether they are positive or negative. We see that while $1-\eta_1$ is always positive, $1-\eta_2$ can change sign depending on the value of $p$. As a result, one needs to know the range in which the value of $p$ of the original depolarisation channel lies, in order to infer which optimal measurement to apply.

\section{The $n^{\text{th}}$-order superswitch}
We now define higher-order superswitches and show that they consist of all commutators and anticommutators of all pairs of commutators and anticommutators generated by the superswitch of order $n-1$, which consists of commutators and anticommutators generated by the one of order $n-2$ and so on. It is instructive to re-derive the zeroth- and first-order superswitches before we show the general case. Given two channels $\E$ and $\F$ with Kraus operators $E_i$ and $F_j$, respectively, the ordinary quantum switch is defined as the channel with Kraus operators
$K^{(0)}_{ij}=E_i F_j \otimes \ketbra{0}_1+ F_j E_i \otimes \ketbra{1}_1$, where the subscript 1 indicates the control ancilla. The resulting supermap is explicitly
\begin{align}
	\Sw^{(0)}_{\omega}(\E,\F)(\rho) = \sum_{i,j} K^{(0)}_{ij} \rho \otimes \omega K^{(0)}_{ij}  	=\frac{1}{4} \sum_{i,j} \left\{E_i, F_j \right\}\rho\left\{E_i, F_j \right\}^\dagger\otimes \omega +  \frac{1}{4} \sum_{i,j} \left[E_i, F_j\right]\rho \left[E_i, F_j\right] ^\dagger \otimes Z \omega Z \,. \label{eq: zero order ss- coms and anticoms}
\end{align}
If we let $\omega=\ketbra{+}$, we obtain
\begin{align}
	\Sw^{(0)}_{\omega}(\E,\F)(\rho) =\frac{1}{4} \sum_{i,j} \left\{E_i, F_j \right\}\rho\left\{E_i, F_j \right\}^\dagger\otimes \ketbra{+}_1 +  \frac{1}{4} \sum_{i,j} \left[E_i, F_j\right]\rho \left[E_i, F_j\right] ^\dagger \otimes \ketbra{-}_1 \,,
\end{align} 

 The first-order superswitch is defined similarly: given four channels $\E$, $\F$, $\EE$ and $\FF$ with Kraus operators $E_i$, $F_j$, $\tilde{E_k}$ and $\tilde{F_l}$ respectively, we first define two quantum switches (i.e.~two zeroth-order superswitches) with Kraus operators
\begin{align}
	K^{(0)}_{ij}&=E_i F_j \otimes \ketbra{0}_1+ F_j E_i \otimes \ketbra{1}_1 \notag \\
	\kk^{(0)}_{ij}&=\ee_i \ff_j \otimes \ketbra{0}_2+ \ff_j \ee_i \otimes \ketbra{1}_2 \,,
\end{align}
with subscripts differentiating between the different control qubits that control the ordering of the channels $\E, \F$ and $\EE, \FF$, respectively. We subsequently define the first-order superswitch as
\begin{align}
	K^{(1)}_{ijkl}&=K^{(0)}_{ij} \kk^{(0)}_{kl} \otimes \ketbra{0}_3+ \kk^{(0)}_{kl} K^{(0)}_{ij} \otimes \ketbra{1}_3 \notag \\
	&= \frac{1}{2} \left\{K^{(0)}_{ij},\kk^{(0)}_{kl}\right\}\otimes \Id+ \frac{1}{2}\left[\kk^{(0)}_{kl},K^{(0)}_{ij}\right]\otimes Z,
\end{align}
which leads to the definition derived in Eq.\@ \eqref{eq: kraus 1st order superswitch}. By induction, the $n^{\text{th}}$-order superswitch will have Kraus operators,
\begin{align}
	K^{(n)}_{i_{n-1}i^\prime_{n-1}}&=K^{(n-1)}_{i_{n-1}} \kk^{(n-1)}_{i^\prime_{n-1}} \otimes \ketbra{0}_{2^n-1}+ \kk^{(n-1)}_{i^\prime_{n-1}} K^{(0)}_{i_{n-1}} \otimes \ketbra{1}_{2^n-1} \notag \\
	&= \frac{1}{2} \left\{K^{(n-1)}_{i_{n-1}}, \kk^{(n-1)}_{i^\prime_{n-1}} \right\}\otimes\Id + \frac{1}{2}\left[K^{(n-1)}_{i_{n-1}}, \kk^{(n-1)}_{i^\prime_{n-1}}\right]\otimes Z,
\end{align}
where the index at the control qubit denotes that at order $n$ there are $2^n-1$ control qubits in total. The symbol $i_{n-1}$ is shorthand for the products of indices $i_1 i_2 \cdots i_{2^n-1}$ associated with the Kraus operators of the $(n-1)^\textrm{th}$-order superswitch. The $n^\textrm{th}$-order superswitch is defined as the channel
\begin{align}
	\Sw_{\omega}^{(n)} &=\frac{1}{4} \sum \left\{K^{(n-1)}_{i_{n-1}}, \kk^{(n-1)}_{i^\prime_{n-1}}   \right\}\rho \left\{K^{(n-1)}_{i_{n-1}}, \kk^{(n-1)}_{i^\prime_{n-1}}  \right\}^\dagger \otimes \omega \notag \\
	&+  \frac{1}{4} \sum \left[K^{(n-1)}_{i_{n-1}}, \kk^{(n-1)}_{i^\prime_{n-1}} \right] \rho \left[K^{(n-1)}_{i_{n-1}}, \kk^{(n-1)}_{i^\prime_{n-1}} \right]^\dagger \otimes Z \omega Z\,. \notag \\
\end{align}
Expanding the expression we see that all possible nested commutators and anticommutator terms are generated. It is worth mentioning that the problem quickly becomes intractable in the general case as it scales doubly exponentially. Specifically, assuming that each channel has $k\leq4$ Kraus operators, the number of possible terms grows as $k^{2^n}2^{2^n-1}$. Also, we note that at order $n$ there are $2^{2^{n+1}-1}$ channels that can be separated in the superswitch by measurements on the control qubits. For instance, the third-order superswitch can separate 32,768 channels, while the fourth has 2,147,483,648. In practice, however, many of the channels at each order end up being the same and thus tracking them becomes less of a daunting task. In any case, for multiple uses of the same channel $\E$ we can write the generic form as follows. If we concisely denote the anticommutators and commutators as $[a,b]_+ = \{a,b\}$ and $[a,b]_- = [a,b]$, we obtain 
\begin{align}
	&\Sw^{(n)}_{\omega_1\cdots \omega_{2^n-1}}(\E,\F\cdots)= \notag \\
	& \,\,\,\, \frac{1}{2^{2^{n+2}-2}} \sum _{i_n}\sum_{a_1,\cdots,a_{2^n-1}=\pm}\left(\left[\cdots,\left[\left[E_i,F_j\right]_{a_1},\left[\EE_i,\FF_j\right]_{a_2}\right]_{a_3}\cdots\right]_{a_{2^n-1}} \rho \left[\cdots,\left[\left[E_i,F_j\right]_{a_1},\left[\EE_i,\FF_j\right]_{a_2}\right]_{a_3}\cdots\right]_{a_{2^n-1}} \right. \notag \\
	&\hspace{50mm}\otimes P_{a_1} \omega_1 P_{a_1} \otimes \cdots \otimes P_{a_{2^n-1}} \omega_{2^n-1} P_{a_{2^n-1}} \Bigg) \,, \label{eq: general switch}
\end{align}
with $P_+=\Id$ and $P_-=Z$.

\subsection{The update rule and recurrence relation}
Even though the nested commutator and anticommutator terms in Eq.\@ \eqref{eq: general switch} appear complicated, in reality one can setup an update rule and evaluate higher-order terms from the previous ones. This is due to the fact that for Pauli channels, the resulting channel at each order is again a Pauli channel, which follows after evaluating commutators and anticommutators of all Kraus operators. By representing a Pauli channel $\E=p_0 \rho + p_1 X\rho X+ p_2 Y\rho Y+ p_3 Z\rho Z$ with its probability vector $\vec{r}=\{p_0, p_1,p_2,p_3\}$, and noting that each term in the sum in Eq.\@ \eqref{eq: general switch} is of the form $[\cdots]_a \rho [\cdots]_a$, it suffices to provide an update rule for the commutator and anticommutator terms for two generic Pauli channels with probability vectors $\vec{r}_i=\{\alpha_i,\beta_i,\gamma_i,\delta_i\}$ and $i=1,2$. We find the update rules (see Appendix E)
\begin{align}
	\acm(\vec{r}_1, \vec{r}_2) &= \{\alpha_1 \alpha_2 +\beta_1\beta_2+\gamma_1 \gamma_2 +\delta_1 \delta_2 \,, \alpha_1 \beta_2+\beta_1 \alpha_2\,, \alpha_1 \gamma_2+\gamma_1 \alpha_2 \,,\alpha_1 \delta_2+\delta_1 \alpha_2\} \,, \notag \\
	\cm (\vec{r}_1, \vec{r}_2) &= \{0, \beta_1 \gamma_2+\gamma_1 \beta_2, \gamma_1 \delta_2+\delta_1 \gamma_2, \delta_1 \beta_2+\beta_1 \delta_2 \}\,, \label{eq: coms and acoms}
\end{align}
where with $\acm(\vec{r}_1, \vec{r}_2)$ we denote the anticommutator term for two channels with probability vectors $\vec{r}_1$ and $\vec{r}_2$, while with $\cm(\vec{r}_1, \vec{r}_2)$ the commutator. Note that the above expressions do not give a channel but a channel multiplied by a probability. The probabilities of occurrence are
\begin{align}
	\Pr(\acm(\vec{r}_1, \vec{r}_2)) &=1 -\Pr(\cm(\vec{r}_1, \vec{r}_2)) \,, \notag \\
	\Pr(\cm(\vec{r}_1, \vec{r}_2)) &= \beta_1 \gamma_2+\gamma_1 \beta_2+ \gamma_1 \delta_2+\delta_1 \gamma_2+ \delta_1 \beta_2+\beta_1 \delta_2 \,. \label{eq: update rule probs}
\end{align}
Eq.\@ \eqref{eq: coms and acoms} can be suggestively rewritten as
\begin{align}
	\acm(\vec{r}_1, \vec{r}_2) &= \Pr(\acm(\vec{r}_1, \vec{r}_2)) \left(  \frac{\{\alpha_1 \alpha_2 +\beta_1\beta_2+\gamma_1 \gamma_2 +\delta_1 \delta_2 \,, \alpha_1 \beta_2+\beta_1 \alpha_2\,, \alpha_1 \gamma_2+\gamma_1 \alpha_2 \,,\alpha_1 \delta_2+\delta_1 \alpha_2\}}{\Pr(\acm(\vec{r}_1, \vec{r}_2))} \right) \,, \notag \\
	\cm (\vec{r}_1, \vec{r}_2) &= \Pr(\cm(\vec{r}_1, \vec{r}_2)) \left( \frac{\{0, \beta_1 \gamma_2+\gamma_1 \beta_2, \gamma_1 \delta_2+\delta_1 \gamma_2, \delta_1 \beta_2+\beta_1 \delta_2 \}}{\Pr(\cm(\vec{r}_1, \vec{r}_2))} \right)\,, \label{eq: update rule}
\end{align}
where now we have both terms in the form of a probability multiplying a probability vector, defining a channel.

If we further assume that all input channels are the same, that is, $\E=\F=\cdots =p_0 \rho + p_1 X\rho X+ p_2 Y\rho Y+ p_3 Z\rho Z$ and represent that single input channel with the probability vector $\vec{r}=\{p_0, p_1,p_2,p_3\}$, this fixes the initial condition for the recurrence relations. In this case, for a single iteration we obtain the channels for the quantum switch, Eq.\@ \eqref{eq: C_+ and C_- channels}, multiplied with the respective probabilities. The procedure can be repeated to obtain the desired higher-order superswitch. 

\subsection{Special instances of the depolarisation channel}

As a first example, we examine special instances of the depolarisation channel which have the special property that for the superswitches of any order, all possible channels after measurements on the ancillas are channels that are mapped to themselves by the commutator and anticommutator terms. It turns out that the only channels with this property are either (i) the channel $D_{\star}(\rho)$,  with the value $p=p_\star=2(1-1/\sqrt{3})$, (ii) the channel $D_{\nicefrac{4}{3}}$, or (iii) the identity map, $\id$. To see this, we first note that the update rule, Eqs.\@ \eqref{eq: update rule probs} and \eqref{eq: update rule}, in the case of two channels with probability vectors of the form $\vec{r}=\{\alpha, \beta,\beta,\beta\}$ and $\vec{v}=\{\alpha^\prime,\beta^\prime,\beta^\prime,\beta^\prime\}$ becomes
\begin{align}
	\Pr(\acm(\vec{r},\vec{v})) =1-6\beta \beta^\prime \,, \, \, 	\Pr(\cm(\vec{r},\vec{v})) = 6\beta \beta^\prime \,,
\end{align}
and
\begin{align}
	\acm(\vec{r}, \vec{v}) &= (1-6\beta \beta^\prime)  \left\{\frac{\alpha \alpha^\prime +3\beta\beta^\prime}{1-6\beta \beta^\prime},\frac{\alpha \beta^\prime +\beta\alpha^\prime}{1-6\beta \beta^\prime},\frac{\alpha \beta^\prime +\beta\alpha^\prime}{1-6\beta \beta^\prime},\frac{\alpha \beta^\prime +\beta\alpha^\prime}{1-6\beta \beta^\prime}\right\} \,, \notag \\
	\cm (\vec{r}, \vec{v}) &= 6\beta \beta^\prime  \left\{0,\frac{1}{3},\frac{1}{3},\frac{1}{3}\right\} \,. \label{eq: update rule dep}
\end{align}
Considering the case of two copies of a depolarisation channel with depolarisation strength $p$, i.e.~$\beta=\beta^\prime=\nicefrac{p}{4}$, and imposing the condition that the anticommutator term, $\acm(\vec{r}, \vec{v})$, maps the channels to themselves, we obtain the equation
\begin{align}
	\frac{p}{4}=\frac{\alpha \beta^\prime +\beta\alpha^\prime}{1-6\beta \beta^\prime} = \frac{\frac{p}{2}-\frac{3p^2}{8}}{1-\frac{3p^2}{8}}\,,
\end{align}
with the only possible solution in the range $p\in[0,\nicefrac{4}{3}]$ being $p=2(1-1/\sqrt{3})$. 
On the other hand, if we impose the same condition for the commutator, we obtain the unique solution $p=\nicefrac{4}{3}$. Along with the trivial case of the identity map, $\id$, there are no other channels that are mapped to themselves by either the commutator or anticommutator term.

Starting with any of these channels, due to the property that they are mapped to themselves, it follows that an $n^{\text{th}}$-order superswitch after measurements on the ancillas will either lead to one of the three aforementioned channels, $\D_{\star} \,, \D_{\nicefrac{3}{4}},  \id $, and thus on average we will have a channel of the form $\alpha_n \D_{\star} +\beta_n \D_{\nicefrac{3}{4}} +\gamma_n  \id $, with $\alpha_n +\beta_n +\gamma_n=1$. Employing the update rule for all nine possible combinations of channels in the sum, we can evaluate the $(n+1)^{\text{th}}$-order superswitch, which, again on average, will be the channel $(\alpha_n^2 (1-c^2)+2\alpha_n \beta_n (1-d)+2\alpha_n \gamma_n) \D_{\star} +(\alpha_n^2 c+2\alpha_n \beta_n d =\frac{2}{3}\beta_n^2 +2\beta_n \gamma_n) \D_{\nicefrac{3}{4}} +(\frac{\beta_n}{3}+\gamma_n^2)  \id$, where we have defined $c=6\beta^2$ and $d=2 \beta$. Thus, we can construct recurrence relations for the coefficients $\alpha_{n+1}, \beta_{n+1}, \gamma_{n+1}$ of the superswitches at an order $n+1$:
\begin{align}
	\alpha_{n+1} &= \alpha_n^2 (1-c^2)+2\alpha_n \beta_n (1-d)+2\alpha_n \gamma_n \notag \\
	\beta_{n+1} &= \alpha_n^2 c+2\alpha_n \beta_n d =\frac{2}{3}\beta_n^2 +2\beta_n \gamma_n \notag \\
	\gamma_{n+1} &= \frac{\beta_n}{3}+\gamma_n^2 \,.
\end{align} 
We note that at order $n=0$ we just have the quantum switch, which leads to the the initial conditions $\alpha_0 = (1-c)\,, \beta_0 = c , \gamma_0=0$ for the recurrence relations.
Even though we do not provide the general solutions, we look for the stationary points,
\begin{align}
	\alpha_{n+1}=\alpha_n \,, \beta_{n+1}= \beta_n \,, \gamma_{n+1}=\gamma_n \,,
\end{align}
which are found to be the triples
\begin{align}
(\alpha_s ,\beta_s, \gamma_s ) : \left\{\left(0,0,1\right) \,,  \left(\frac{1}{2}, \frac{1}{2}+\frac{1}{4}(\sqrt{3}-2),\frac{1}{4}(2-\sqrt{3})\right)\,,\left(0, \nicefrac{3}{4},\nicefrac{1}{4}\right) \right\}\,,
\end{align}
which correspond to the channels $\id$, $\frac{1}{2} \D_{\star}+\left(\frac{1}{2}+\frac{1}{4}(\sqrt{3}-2)\right)\D_{\nicefrac{4}{3}}+\left(\frac{1}{4}(2-\sqrt{3})\right) \id$, and $\frac{3}{4}\D_{\nicefrac{4}{3}}+\frac{1}{4} \id$,  respectively. 

The first solution corresponds to the case where we start with the identity channel, $\id$, which is a trivial case.
The second solution corresponds to the case of the channel $\D_\star$, and gives the optimal guessing in the limit of $n\rightarrow\infty$. Explicit evaluation leads to a guessing probability in the limit
\begin{align}
	\pg^{(\star,\infty)} = \frac{1}{12}\left(6+\sqrt{3}\right) \approx 0.644 \,,
\end{align}
that upper bounds the guessing probability of any $n$-order superswitch. 
This value should be contrasted with the guessing probability 
\begin{align}
	\pg^{(\star)} = \frac{1}{\sqrt{3}} \approx 0.577 \,,
\end{align}
that is achieved by sending the states directly through the channel $\D_\star$ without the use of a switch, corresponding to at most an 11.6\% improvement.  In this case, each higher-order superswitch gives a higher guessing than lower-order ones. However, we note that the first few orders quickly converge to the upper bound and thus minimal improvement is achieved with each subsequent superswitch. We evaluate the first few orders of superswitches and we find that the quantum switch achieves a guessing of 0.601, while the next four-order superswitches achieve 0.619, 0.631, 0.636, 0.639, respectively.

The final solution is obtained if we start from copies of the channel $\D_{\nicefrac{4}{3}}$. Once again the solution gives the limiting value of the guessing probability of superswitches, which is found to be
\begin{align}
	\pg^{(\nicefrac{4}{3},\infty)} = \frac{3}{4} =0.75 \,,
\end{align}
which should be contrasted with the guessing probability 
\begin{align}
	\pg^{(\nicefrac{4}{3})} = \frac{2}{\sqrt{3}} \approx 0.667 \,,
\end{align}
that is achieved by sending the states directly through the channel $\D_{\nicefrac{4}{3}}$ without the use of a switch, corresponding to a 12.5\% improvement. Interestingly, in this case the limiting value does not upper bound all superswitches and the highest guessing is achieved by the quantum switch with a guessing of $\nicefrac{7}{9} \approx 0.778$, which is an 16.7\% improvement, with each subsequent higher-order switch achieving a lower guessing probability, with the values tending to the limit $\nicefrac{3}{4}$. The first four superswitches achieve the guessing probabilities 0.778, 0.753, 0.75004, 0.750000006, 0.7500000000000001, and thus we see that they rapidly converge to the (sub-optimal) limiting value.

\subsection{State discrimination with higher-order superswitches}
We now study higher-order superswitches for the following two channels: (i) the depolarisation channel and (ii) a Pauli channel with $p_1=0$.

\subsubsection{The depolarisation channel}
We evaluate the superswitches of up to order four and derive the expressions for the guessing probabilities, where we assume that after measurements on the ancillas, Bob always applies an appropriate optimal measurement for the channel corresponding to the obtained outcomes, communicated to him by the communication provider. Since the expressions are long we only mention them in Appendix F.

In Fig. \ref{fig: depolarisation - superswitches} we plot the guessing probabilities for the depolarisation channel and superswitches up to order four. We find that there are regions in the parameter $p$ with very different behaviours and an intricate picture emerges: there exist regions where the higher-order superswitches outperform the lower-order superswitches, regions where the superswitches do progressively worse than the channel, as well as regions in which some switches perform better than others. Explicitly, in the region to the left of the dashed orange line the higher the order of the superswitch, the worse the guessing probability becomes. It is clear that it is not beneficial to employ a superswitch protocol in this region. Similarly, in the region between the dashed orange line and the dotted black line, the superswitches perform worse than the channel, but comparing the superswitches in terms of performance, the ordering is not clear throughout the region. The region between the black dotted line and the leftmost dashed purple line is interesting as for some values of $p$ some of the switches may perform better than the channel but not all. For instance, for the value $p\approx 0.75$, the quantum switch performs worse than the channel, the first-order superswitch performs better than the quantum switch but still worse than the channel, and it is only when we get to the second-order switch that we start seeing an advantage over the guessing probability of the channel. In the region between the two dashed purple lines all superswitches outperform the channel and their ordering follows the pattern that the higher the order of the superswitch the higher the guessing probability. We expect that this pattern holds for all the $n$\ts{th}-order superswitches in this region but we have no proof of this claim. 

Perhaps the most interesting behaviours occur for values of $p>1$. In this interval, there exist regions where the ordering of the superswitches in terms of guessing probabilities does not follow a clear pattern. For example, for the value $p\approx 1.05$ the quantum switch has a significant advantage over the channel, which, however, is reduced for the first-order superswitch, the second-order superswitch outperforms both, while the third-order performs worse than the second but better than the previous ones; finally, the fourth-order superswitch performs the best. We note that in the case $p=1$ where the channel becomes completely depolarising and the best guessing can not exceed uniform guessing, all superswitches perform significantly better with a guessing that is at least $\nicefrac{5}{8}=0.65$, an increase of at least 30\%. Interestingly, for $p=1$ the quantum switch and first-order superswitch have the same guessing probability, while higher-order superswitches exceed both. In the region between the rightmost purple and orange dashed lines, we once again find some subtlety in the behaviour of the superswitches, while at the right of the orange dashed line, we find a region where higher-order switches perform worse than lower-orders. However, in contrast to the leftmost region, in this region all superswitches outperform the channel. From the above considerations, we conjecture that for the depolarisation channel, given accessibility up to the $n^{\text{th}}$-order superswitch, the guessing probability can be split into three regions: (i) a region where the channel performs better than any of the switches, (ii) a region where the $n^{\text{th}}$-order switch performs the best, and (iii) a region where the quantum switch performs the best.

\begin{figure}[!t]
	\centering	\includegraphics[width=0.7\linewidth]{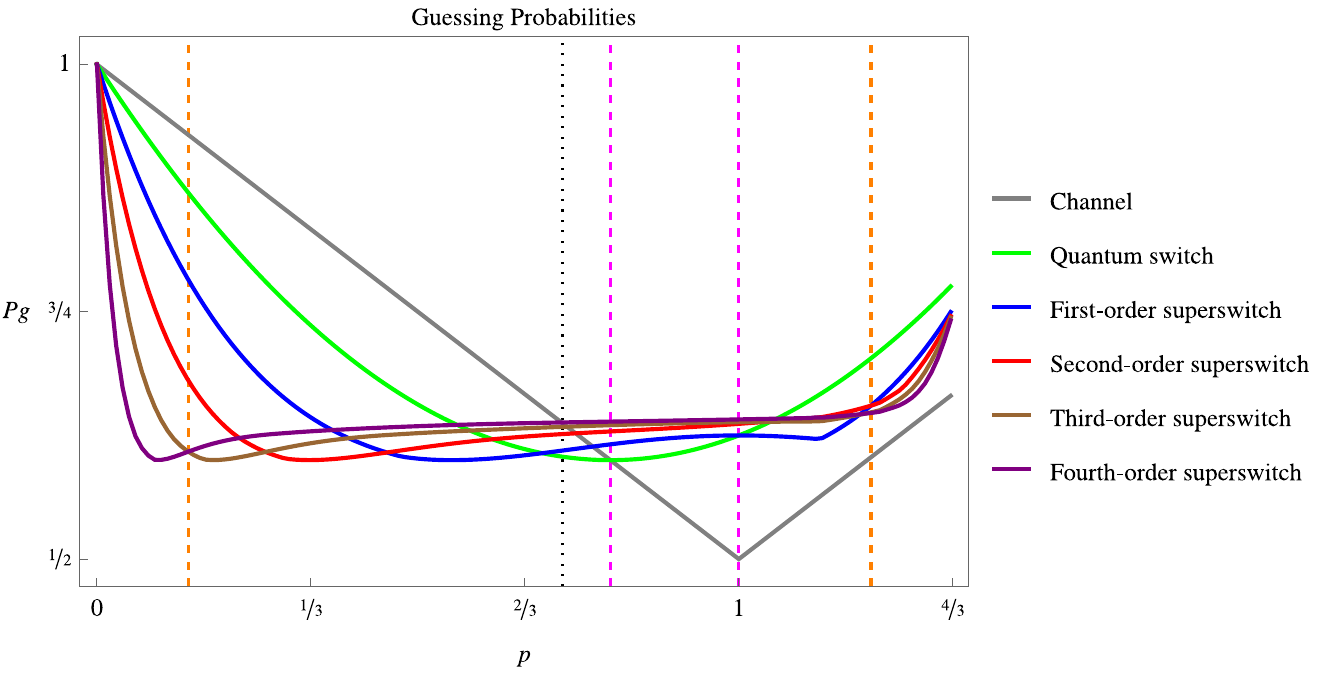}
	\caption{Plot of the guessing probabilities for the orthogonal pair of states $\ketbra{0}, \ketbra{1}$, occurring with equal \emph{a priori} probabilities $\nicefrac{1}{2}$, subjected to depolarisation noise, as well as the superswitches.}
	\label{fig: depolarisation - superswitches}
\end{figure} 

\begin{figure}[!b]
	\centering	\includegraphics[width=0.5\linewidth]{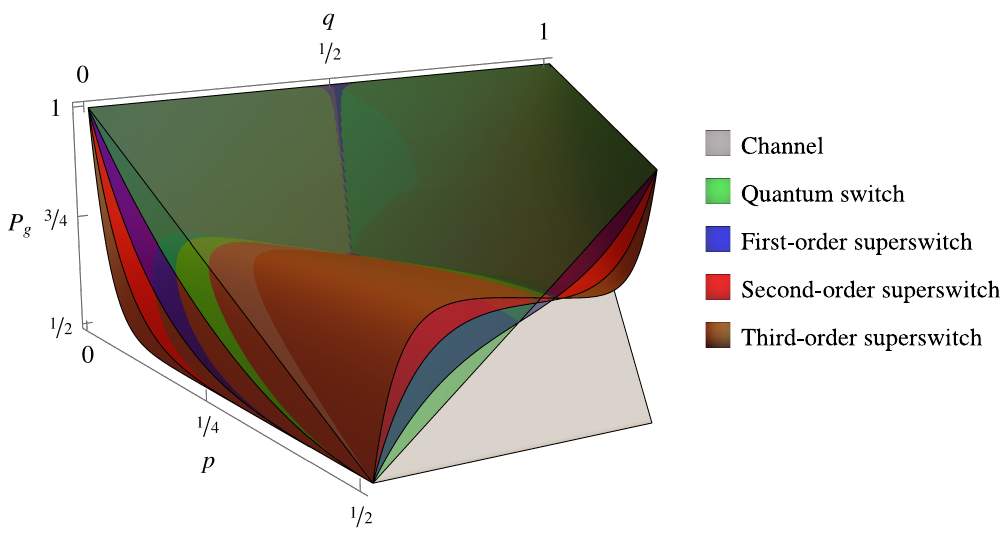}
	\caption{Guessing probabilities for the orthogonal pair of states $\ketbra{0}, \ketbra{1}$, occurring with equal \emph{a priori} probabilities $\nicefrac{1}{2}$, subjected to the channel $Q_{p,q}(\rho)$ (grey), quantum switch (green), first-order superswitch(blue), second-order superswitch (red), and third-order superswitch (brown).}
	\label{fig: Q channel - superswitches}
\end{figure} 

\subsubsection{Pauli channels with $p_1=0$}

\begin{figure}[!t]
	\centering	\includegraphics[width=0.9\linewidth]{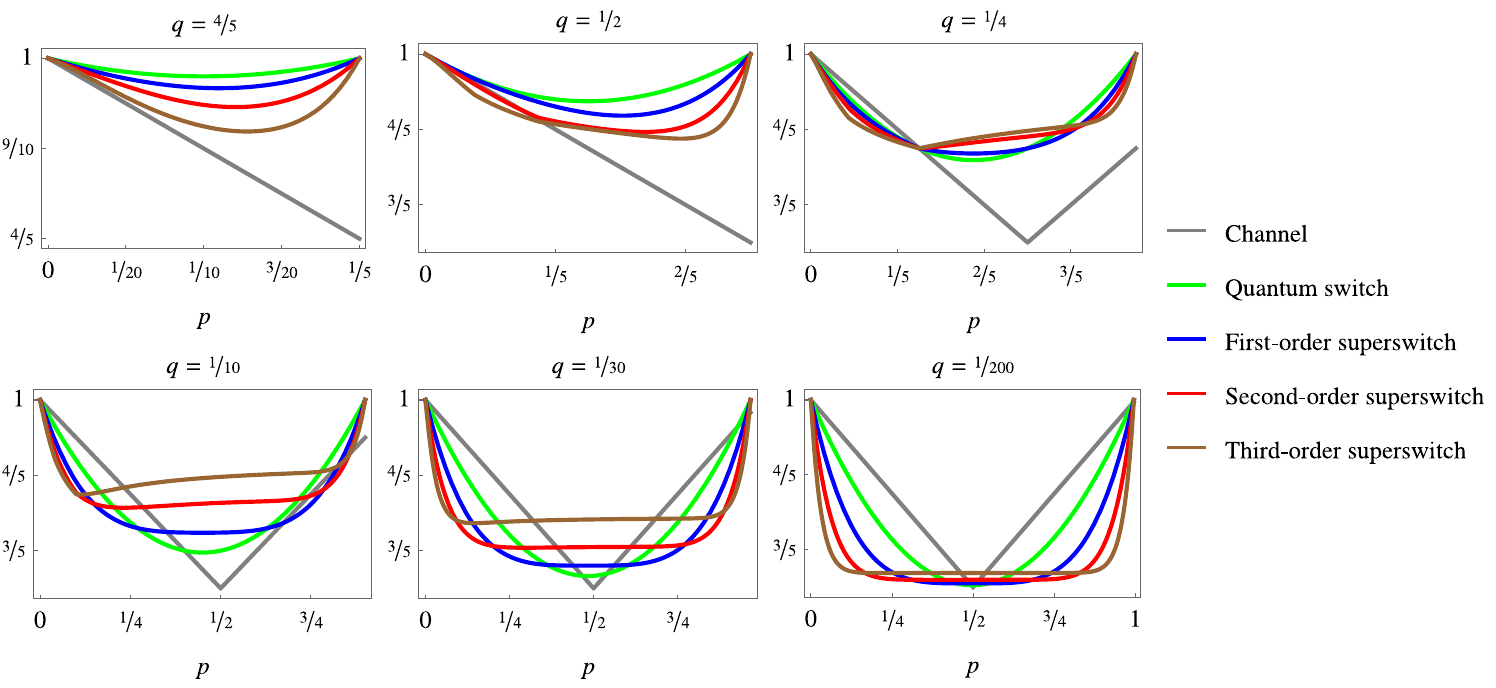}
	\caption{Guessing probabilities for the orthogonal pair of states $\ket{0}, \ket{1}$, occurring with equal \emph{a priori} probabilities $\nicefrac{1}{2}$, subjected to the channel $Q_{p,q}(\rho)$ for various values of the parameter $q$ as a function of $p$. }
	\label{fig: Q channel - superswitches -grid}
\end{figure} 

Another interesting case occurs for a channel where one of the $p_i$'s is equal to 0, for instance  the channel
 \begin{align}
 	Q_{p,q}(\rho)=(1-p-q)\rho + p Y\rho Y+q Z\rho Z \,. \label{eq: Q channel}
 \end{align} 
 Similarly to the depolarisation channel, we find that intricate behaviour arises in the parameter space $(p,q)$ of the channel, with regions with very different ordering between the guessing probabilities of the channel and the superswitches. In Fig.\@ \ref{fig: Q channel - superswitches} we consider the orthogonal pair of states $\ket{0},\ket{1}$ with equal \emph{a priori} probabilities $\nicefrac{1}{2}$ and plot the guessing probabilities of the channel and the superswitches up to order 3, as functions of the parameters $q$ and $p$ of the channel.

In Fig.\@ \ref{fig: Q channel - superswitches -grid} we show sections of the above 3D plot for 6 different fixed values of the parameter $q$. Specifically, for large values of $q$, the channel gives the worst guessing for all values of $p$, however, the superswitches perform worse the larger their order is and thus the quantum switch performs optimally in such cases. For the value $q=\nicefrac{1}{2}$, some of the superswitches perform even worse than the channel. For the value $q=\nicefrac{1}{4}$ a behaviour similar to the case of the depolarisation channels emerges with three special regions where either the channel, highest order switch or quantum switch dominates. With increasing values of $q$ the differences in performance become more pronounced and as $q$ tends to 0, the channel tends to a phase-flip channel where no advantage can be obtained through indefinite causal order, similar to the bit-phase flip channel examined earlier. 

\begin{figure}[!b]
	\centering	\includegraphics[width=0.65\linewidth]{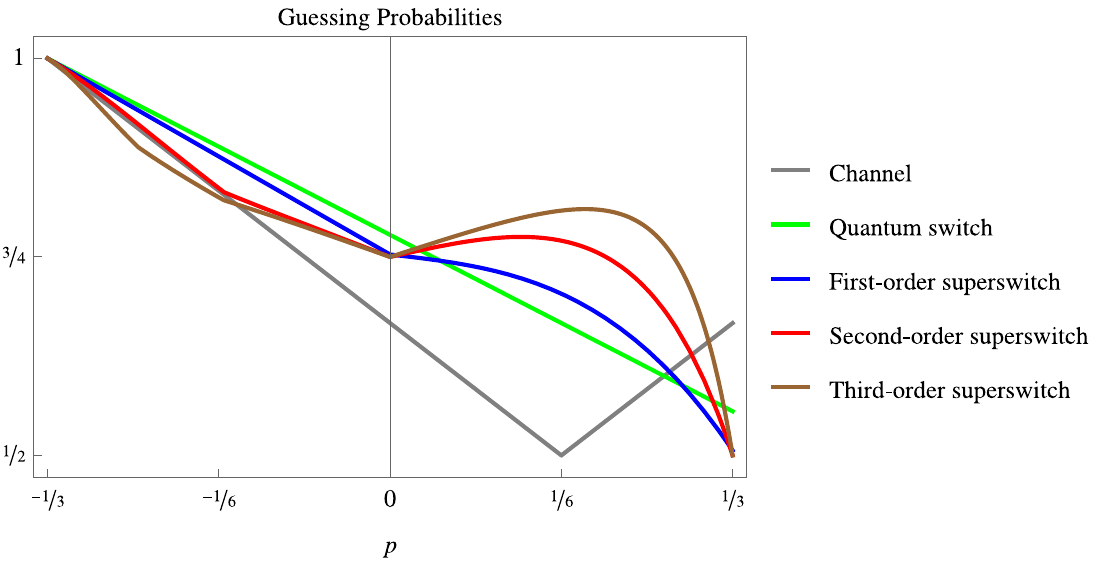}
	\caption{Guessing probabilities for the orthogonal pair of states $\ket{0}, \ket{1}$, occurring with equal \emph{a priori} probabilities $\nicefrac{1}{2}$. subjected to the channel $\tilde{Q}_{p,q}(\rho)$ (grey), quantum switch (green), first-order superswitch (blue), and second-order superswitch (red), and third-order superswitch (brown).}
	\label{fig: Qtilde channel - superswitches}
\end{figure} 

As another interesting special case, we further investigate an instance of the above channel,
$\tilde{Q}_p (\rho)=\frac{1}{3}\rho + \left(\frac{1}{2}+p\right) Y\rho Y+ \left(\frac{1}{2}-p\right) Z\rho Z$, with $p\in[-\nicefrac{1}{3},\nicefrac{1}{3}]$, which follows after the substitution $p\rightarrow 1/3+p$ and $q\rightarrow 1/2-p$.
The plot of guessing probabilities are shown in Fig.\@ \ref{fig: Qtilde channel - superswitches}.

Again, we find that there are regions with completely different behaviours, but for values $-0.14<p<0.26$, all superswitches outperform the channel. However, their respective ordering depends on the values of $p$.

\section{Superswitches for general Pauli channels}

\begin{figure}[!b]
\begin{tabular}{ccc}
	\includegraphics[width=0.31\linewidth]{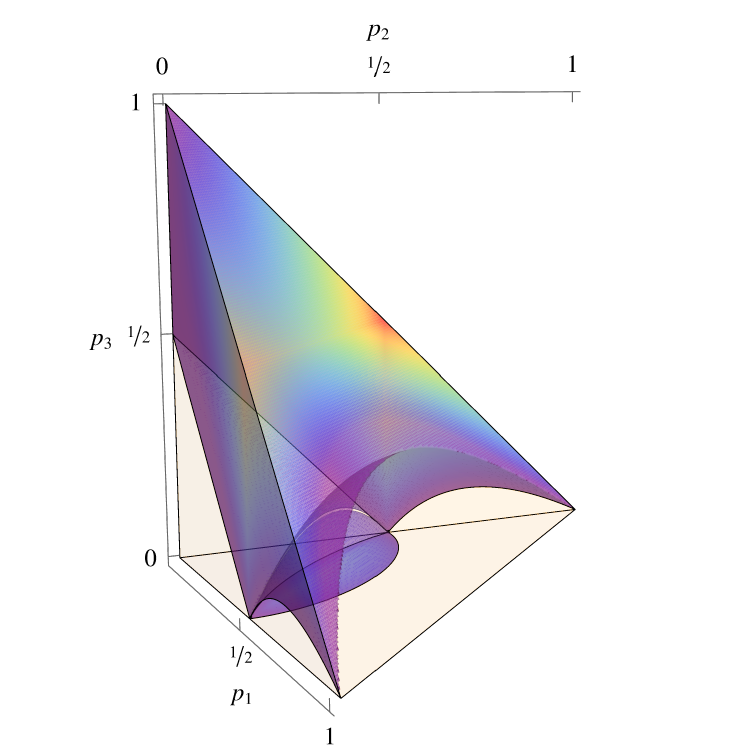} &   \includegraphics[width=0.31\linewidth]{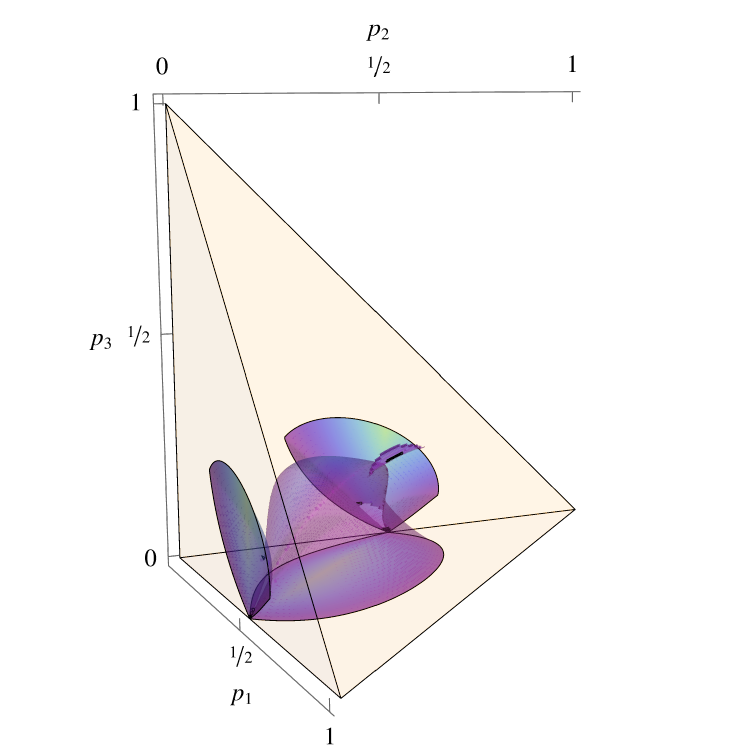} & \includegraphics[width=0.36\linewidth]{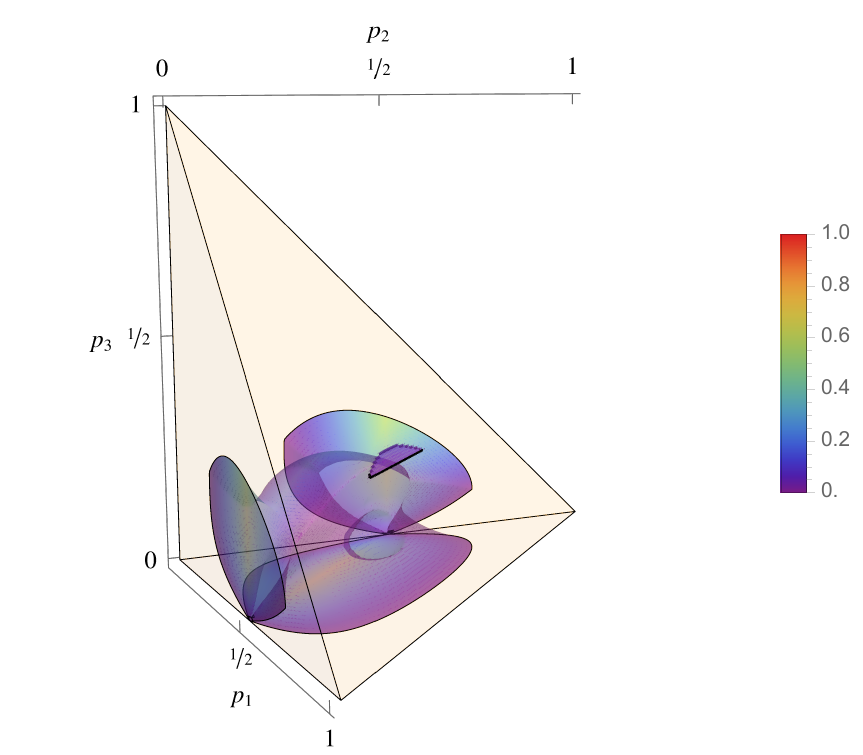} \\  
	(a) & (b) & (c)\\[6pt]
\end{tabular}
	\caption{(a) The region $\mathcal{R}^{(\Sw)}$ where a higher guessing is achieved by the quantum switch over a general Pauli channel.  (b) The region $\mathcal{R}^{(\mathcal{SS}_1)}$ where the first-order superswitch achieves a higher guessing over both the channel and the quantum switch. (c) The region $\mathcal{R}^{(\mathcal{SS}_2)}$ where the second-order superswitch has higher guessing in comparison to that of the channel and all previous-order superswitches, for the two orthogonal states $\ketbra{0},\ketbra{1}$ with equal \emph{a priori} probabilities. The graphics are colour-coded to show the increase in guessing probability in relation to that of the channel.}
	\label{fig: general Pauli vs switches}
\end{figure}

Here we extend the analysis to the set of all Pauli channels and derive general statements on the performance of superswitches, for an ensemble consisting of two orthogonal states, $\Omega=\{\nicefrac{1}{2},\ketbra{i}\}_{i=0,1}$. Given a general Pauli channel $\E_{\vec{p}} (\rho) = \sum_i p_i \sigma_i \rho \sigma_i$, where  $\vec{p}=(p_0,p_1,p_2,p_3)$, we examine regions in parameter space where a higher guessing probability is achieved with the quantum switch compared to the channel. Specifically, the region corresponding to the convex set of all Pauli channels, $\mathcal{R}^{(\E)}$, is defined as
\begin{align}
	\mathcal{R}^{(\E)} = \{  p_1,p_2,p_3 \in [0,1] \, |  \,\,  p_1+p_2+p_3 \leq 1 \} \, \,,
\end{align} 
and forms a tetrahedron with the four vertices $(0,0,0), (1,0,0), (0,1,0), (0,0,1)$ in the parameter space $(p_1,p_2,p_3)$. 
The region of improvement by the quantum switch, $\mathcal{R}^{(\Sw)} $, is defined as
\begin{align}
	\mathcal{R}^{(\Sw)} = \{ p_1,p_2,p_3 \in \mathcal{R}^{(\E)} \, | \,\, \pg^{(\Sw)}>\pg^{(\E)} \} \,,
\end{align}  
where $\pg^{(\E)}$ denotes the guessing probability of a Pauli channel and $\pg^{(\Sw)}$ denotes the guessing probability with the quantum switch protocol. In Fig.\@ \ref{fig: general Pauli vs switches} we show the region of allowed values of $p_1,p_2,p_3$ where the inequality $\pg^{(\Sw)}>\pg^{(\E)}$ holds.

To quantify the set of Pauli channels for which it is advantageous to use the quantum switch protocol, we evaluate the volume of the region $\mathcal{R}^{(\Sw)} $ through the triple integral
\begin{align}
	V_{\Sw} = \int_{\mathcal{R}^{(\mathcal{S})}} \mathrm{d}V_{\mathcal{S}}= \int_{p_1,p_2,p_3 \in \mathcal{R}^{(\Sw)} } \mathrm{d}p_1\, \mathrm{d}p_2\, \mathrm{d}p_3 \,,
\end{align}
and compare to the volume of the tetrahedron that corresponds to the region of all Pauli channels, $\mathcal{R}^{(\E)}$, which has a volume of $V_{\E} = \nicefrac{1}{6}$.
After a numerical evaluation we find the value $V_{\Sw} \approx 0.093$, which gives a ratio of $\nicefrac{V_{\Sw}}{V_{\E}}= 0.555$. It follows that among all Pauli channels, and for the orthogonal pair of states $\ketbra{0},\ketbra{1}$, an improvement is obtained for approximately $55.5\%$ of the channels when the quantum switch is used.

\begin{figure}[!t]
\centering
		\includegraphics[width=0.4\linewidth]{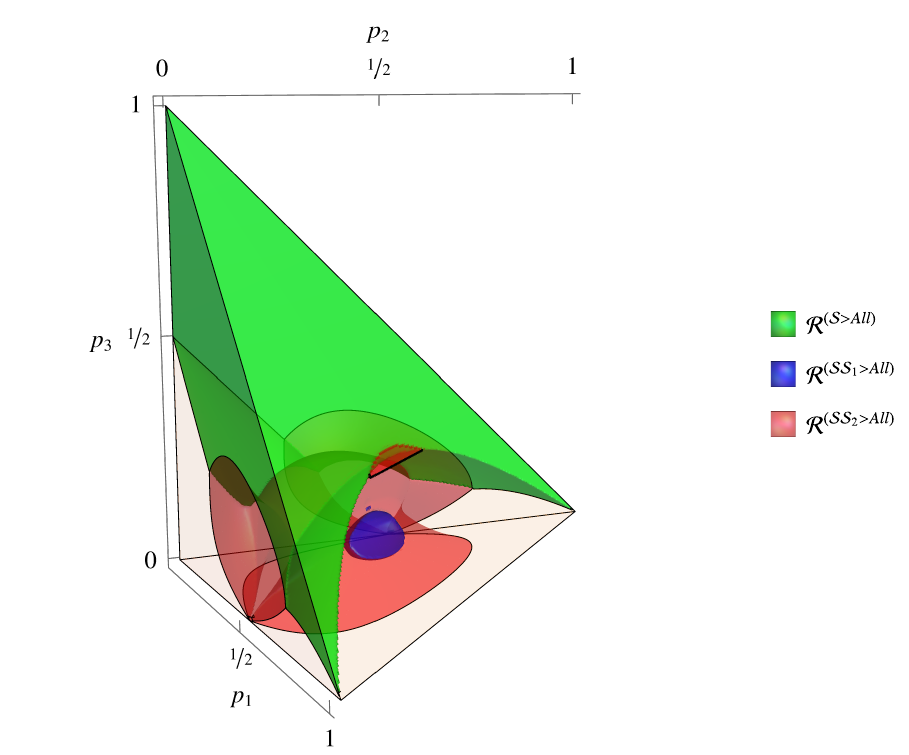} 
	\caption{ The three regions $ \mathcal{R}^{(\Sw>All)}$ (green), $\mathcal{R}^{(\mathcal{SS}_1>All)}$ (blue), and $\mathcal{R}^{(\mathcal{SS}_2>All)}$ (red) where the quantum switch and first and second-order superswitches achieve the highest guessing probability among the channel and all other superswitches, for the two orthogonal states $\ketbra{0},\ketbra{1}$ with equal \emph{a priori} probabilities. }
	\label{fig: superswitches}
\end{figure}

We perform a similar analysis for the first-order superswitch, comparing to the channel and the quantum switch. Denoting by $\pg^{(\mathcal{SS}_1)}$ the guessing probability achieved by the first-order superswitch protocol, in Fig.\@ \eqref{fig: general Pauli vs switches} we also show the region of allowed values of $p_1,p_2,p_3$ where the inequalities $\pg^{(\mathcal{SS}_1)}>\pg^{(\E)}$ and $\pg^{(\mathcal{SS}_1)}>\pg^{(\Sw)}$ simultaneously hold. The region of improvement over the channel, $\mathcal{R}^{(\mathcal{SS}_1)} $, is defined as
\begin{align}
	\mathcal{R}^{(\mathcal{SS}_1)} = \{ p_1,p_2,p_3 \in \mathcal{R}^{(\E)} \, | \,\, \pg^{(\mathcal{SS}_1)}>\pg^{(\E)} \text{ and } \pg^{(\mathcal{SS}_1)}> \pg^{(\Sw)} \} \,,
\end{align} 
and its volume is found to be $V_{\mathcal{SS}_1}  \approx 0.022$, which, when compared to volume of the region of all channels, gives the ratio $\nicefrac{V_{\mathcal{SS}_1}}{V_{\E}}= 0.138$. As a result, the first-order superswitch can further increase the guessing probability for approximately $13.8\%$ of channels.

We also consider the second-order superswitch and we denote by $\pg^{(\mathcal{SS}_2)}$ the guessing probability achieved by the second-order superswitch protocol, and define the region of improvement, $\mathcal{R}^{(\mathcal{SS}_2)}$, over the channel and previous-order superswitches as
\begin{align}
	\mathcal{R}^{(\mathcal{SS}_2)} = \{ p_1,p_2,p_3 \in \mathcal{R}^{(\E)}\, | \,\, \pg^{(\mathcal{SS}_2)}>\pg^{(\E)} \text{ and } \pg^{(\mathcal{SS}_2)}>\pg^{(\Sw)} \text{ and } \pg^{(\mathcal{SS}_2)}>\pg^{(\mathcal{SS}_1)} \} \,, \label{eq: SS2>all}
\end{align} 
we find that it has a volume of $V_{\mathcal{SS}_2}  \approx 0.0335$, which gives the ratio $\nicefrac{V_{\mathcal{SS}_2}}{V_{\E}}= 0.201$. Thus, the second order superswitch can further increase the guessing probability for approximately $20.1\%$ of the channels in the parameter space. 
There is an interesting behaviour in the regions of improvement, with the quantum switch offering an improvement for approximately 55.5\% of all Pauli channels. The range of further improvement by the first-order superswitch  occurs only for 13.8\% of channels, which, interestingly, is superseded by the second-order superswitch that is offering an increase over the channel and all lower-order superswitches for approximately 20.1\% of all Pauli channels. We see that the regions of improvement by superswitches do not follow a simple pattern. 
 These statements alone do not take into consideration the relative performance between the superswitches, as they only quantify the regions of improvement in comparison to the channel and lower-order superswitches. 

Even more interestingly, there are regions where lower-order superswitches can still dominate over higher-order ones.
 To compare the performance of the superswitches, we define the three regions of dominance: (i) the region $\mathcal{R}^{(\Sw>All)}$ where the quantum switch beats both the channel and the first- and second-order superswitches, 
\begin{align}
		\mathcal{R}^{(\Sw>All)} = \{ p_1,p_2,p_3 \in \mathcal{R}^{(\E)} \, | \,\, \pg^{(\Sw)}>\pg^{(\E)} \text{ and } \pg^{(\Sw)}> \pg^{(\mathcal{SS}_1)} \text{ and } \pg^{(\Sw)}> \pg^{(\mathcal{SS}_2)} \} \,,
\end{align} 
(ii) the region $\mathcal{R}^{(\mathcal{SS}_1> All)} $ where the first-order superswitch beats the channel, the quantum switch, and the second-order superswitch,
\begin{align}
	\mathcal{R}^{(\mathcal{SS}_1> All)} = \{ p_1,p_2,p_3 \in \mathcal{R}^{(\E)} \, | \,\, \pg^{(\mathcal{SS}_1)}>\pg^{(\E)} \text{ and } \pg^{(\mathcal{SS}_1)}>\pg^{(\Sw)} \text{ and } \pg^{(\mathcal{SS}_1)}>\pg^{(\mathcal{SS}_2)}\} \,,
\end{align} 
and (iii) the region $\mathcal{R}^{(\mathcal{SS}_2> All)}= \mathcal{R}^{(\mathcal{SS}_2)}$ where the second-order superswitch beats the channel, the quantum switch, and the first-order superswitch, already defined in Eq.\@ \eqref{eq: SS2>all}.

All regions are contrasted in Fig.\@ \ref{fig: superswitches}.  Their volumes are found to be $V_{\mathcal{\Sw}> All}\approx0.0751$, $V_{\mathcal{SS}_1> All}\approx0.0004$, and $V_{\mathcal{SS}_2> All}\approx0.0335$, or as ratios to the volume of the tetrahedron of all Pauli channels, $\nicefrac{V_{\mathcal{\Sw}}}{V_{\E}}= 0.450$, $\nicefrac{V_{\mathcal{SS}_1> All}}{V_{\E}}= 0.003$ and $\nicefrac{V_{\mathcal{SS}_2> All}}{V_{\E}}= 0.201$. The following picture thus emerges for all possible Pauli channels and the ensemble $\Omega$: (i) for 34.6\% of them a protocol with a superswitch of order less than or equal to two cannot increase the guessing probability, (ii) for 45\% the quantum switch achieves the highest guessing probability, (iii) for 0.3\% the first-order superswitch achieves the highest guessing probability, and (iv) for 20.1\% of them the highest guessing probability is achieved with the second-order superswitch.

\section{Two-qubit state Pauli channels}

The defining equation for the superswitches, Eq.\@ \eqref{eq: general switch}, is agnostic to the state space dimension. By an appropriate modification of the update rules, Eqs.\@ \eqref{eq: update rule probs}, \eqref{eq: update rule}, we can extend our results to Pauli channels in any dimension. We will restrict the discussion to dimension four as a proof-of-principle demonstration that the extension to any dimension $2^n$ is straightforward. A Pauli channel in dimension four is of the form
\begin{align}
	\E_{\vec{p}}(\rho) = \sum_{i} \vec{r}_i R_i \rho R_i^\dagger \,,
\end{align}
where 
\begin{align}
	R =	\{&\Id \otimes \Id\,, \,  \Id \otimes X\,, \, \Id \otimes Y\,, \, \Id \otimes Z \,, \,
	X \otimes \Id\,, \,  X \otimes X\,, \, X \otimes Y\,, \, X \otimes Z \,, \,  \notag \\
	&Y \otimes \Id\,, \,  Y \otimes X\,, \, Y \otimes Y\,, \, Y \otimes Z \,, \, Z \otimes \Id\,, \,  Z \otimes X\,, \, Z \otimes Y\,, \, Z \otimes Z\} \,, \label{eq: tensor of Paulis}
\end{align}
collects tensor products of Pauli matrices and $\vec{r}$ is the probability vector with the respective probabilities of occurrence.

To construct the update rule for the superswitches, similar to dimension two, one needs to derive the commutators and anticommutators for two Pauli channels with probability vectors $\vec{r}$ and $\vec{v}$. Since the expressions are long, they are included in Appendix G. Through these, the superswitch channels of any order can be evaluated for a given channel.

With the above definitions we now study superswitches in $d=4$ for  two types of depolarisation channels. The standard definition of a depolarisation channel in $d>2$ follows that of the two-dimensional case as a probabilistic mixing of the state and the maximally mixed state, that is,
\begin{align}
	\D_{p}^{(d)} (\rho) = (1-p) \rho + p \frac{\Id _d}{d}  \,, \text{ with }  0\leq p \leq \frac{d^2}{d^2-1} \,,
\end{align}
where $\Id_d$ denotes the $d$-dimensional identity matrix. 
In this section we focus on the depolarisation channel $\D^{(4)}_p$ and for notational convenience we now drop the superscript ``$(4)$". The depolarisation channel can also be written in the Kraus representation as $\D_s(\rho)=\sum_i \vec{v}_i R_i \rho R_i ^\dagger$, with the probability vector being
\begin{align}
	\vec{v}_i = \left\{\left(1-\frac{15s}{16}\right), \frac{s}{16},\frac{s}{16},\frac{s}{16},\frac{s}{16},\frac{s}{16},\frac{s}{16},\frac{s}{16},\frac{s}{16},\frac{s}{16},\frac{s}{16},\frac{s}{16},\frac{s}{16},\frac{s}{16},\frac{s}{16},\frac{s}{16}\right\} \,,
\end{align}
with $s$ satisfying $0\leq s \leq \nicefrac{16}{15}$.

We also consider a variant of the depolarisation channel, whose definition can be seen as two depolarisation channels with different noise parameters $p$ and $q$, each acting on different parts of a bipartite state $\rho_{AB}$. Specifically, by assuming that there is a probability $p$ that an $X,Y$ or $Z$ error is applied on one part of the state and $1-p$ of no error, and  similarly a probability $q$ that an $X,Y$ or $Z$ error is applied on the other part with $1-q$ the probability of no error, we arrive at the two-parameter channel
\begin{align}
	\Delta_{p,q}(\rho) = (\D_p\otimes\D_q) (\rho_{AB})\,.
\end{align}
Writing $\D_p = (1-p) \id + p \Lambda $, where $\Lambda$ denotes the completely depolarising map, we obtain
\begin{align}
	\Delta_{p,q}(\rho) &=\Big((1-p)(1-q) \id_A \otimes \id_B +(1-p)q \id_A \otimes \Lambda_B +p(1-q) \Lambda_A \otimes \id_B +pq \Lambda_A \otimes \Lambda_B\Big)(\rho_{AB}) \notag \\
	&=(1-p)(1-q) \rho_{AB} +(1-p)q \rho_A \otimes \frac{\Id_2}{2} +p(1-q) \frac{\Id_2}{2}\otimes \rho_B +pq \frac{\Id_4}{4} \,,
\end{align}
where $\Id_d$ denotes the identity in $d$ dimensions, and $\rho_A, \rho_B$ the partial traces of $\rho_{AB}$ with respect to $B$ and $A$, respectively.
By using the identity $4 \Id_2 = \Id_2\rho \Id_2 +X\rho X +Y\rho Y+Z\rho Z$, we can write the channel in the Kraus representation as
\begin{align}
	\Delta_{p,q}(\rho) = \sum_{i} \vec{r}_i R_i \rho R_i^\dagger \,,
\end{align}
where $R$ are the tensor products of Pauli matrices defined in Eq.\@ \eqref{eq: tensor of Paulis}, and 
\begin{align}
	\vec{r} = \Bigg\{&\left(1-\frac{3p}{4}\right)\left(1-\frac{3q}{4}\right), \left(1-\frac{3p}{4}\right) \frac{q}{4},\left(1-\frac{3p}{4}\right) \frac{q}{4},\left(1-\frac{3p}{4}\right) \frac{q}{4},
 \notag \\
	&	\,\, \,\, \frac{p}{4}\left(1-\frac{3q}{4}\right), \frac{pq}{16},\frac{pq}{16},\frac{pq}{16}, \frac{p}{4}\left(1-\frac{3q}{4}\right), \frac{pq}{16},\frac{pq}{16},\frac{pq}{16},\frac{p}{4}\left(1-\frac{3q}{4}\right), \frac{pq}{16},\frac{pq}{16},\frac{pq}{16}	\Bigg\} \,,
\end{align}
are the respective probabilities of each error occurring. In order for the map to be CPTP, the values of $p$ and $q$ are constrained by $0\leq p ,q \leq \nicefrac{4}{3}$.

\begin{figure}[!t]
	\begin{tabular}{ccc}
		\includegraphics[width=0.32\linewidth]{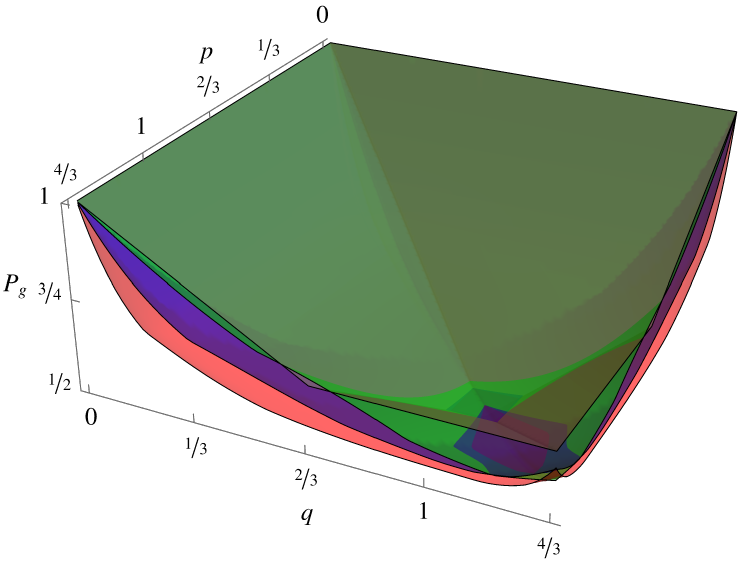} &   \includegraphics[width=0.32\linewidth]{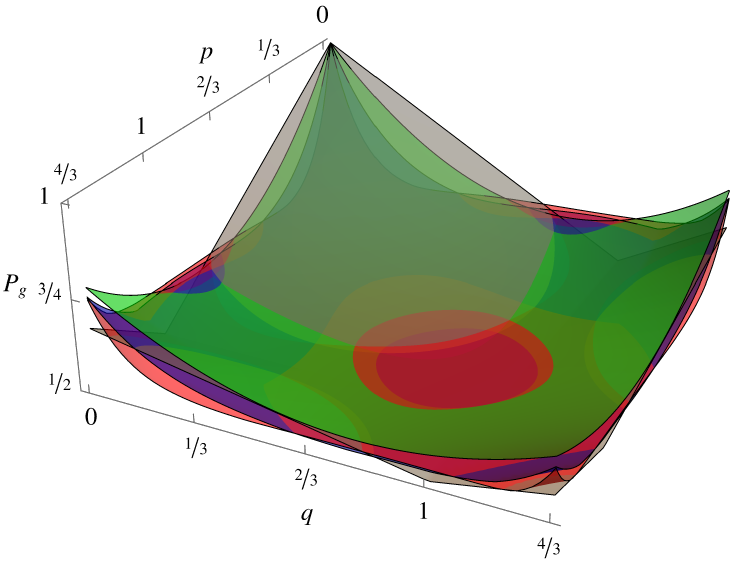} & \includegraphics[width=0.32\linewidth]{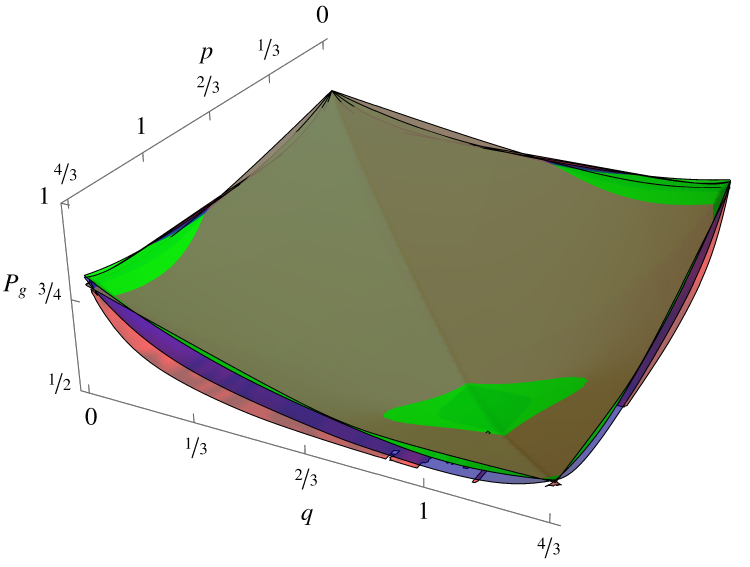}  \\[6pt]
		(a) & (b) & (c) \\ [6pt]
		\includegraphics[width=0.32\linewidth]{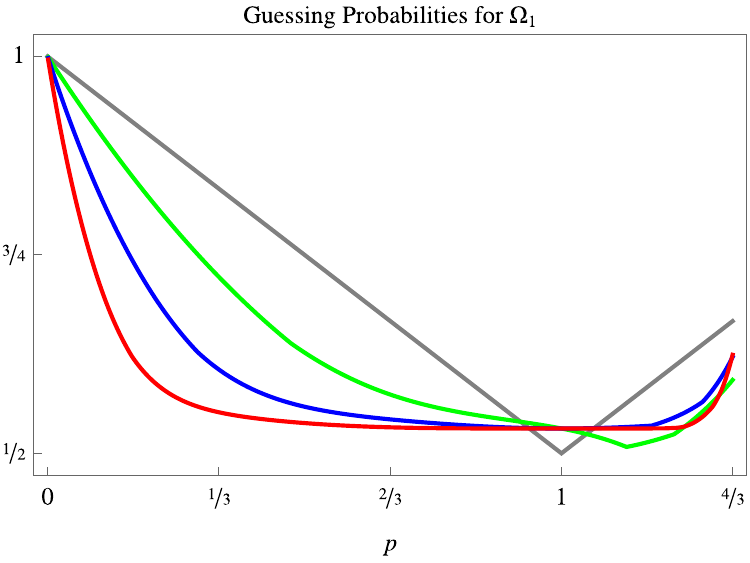} &   \includegraphics[width=0.32\linewidth]{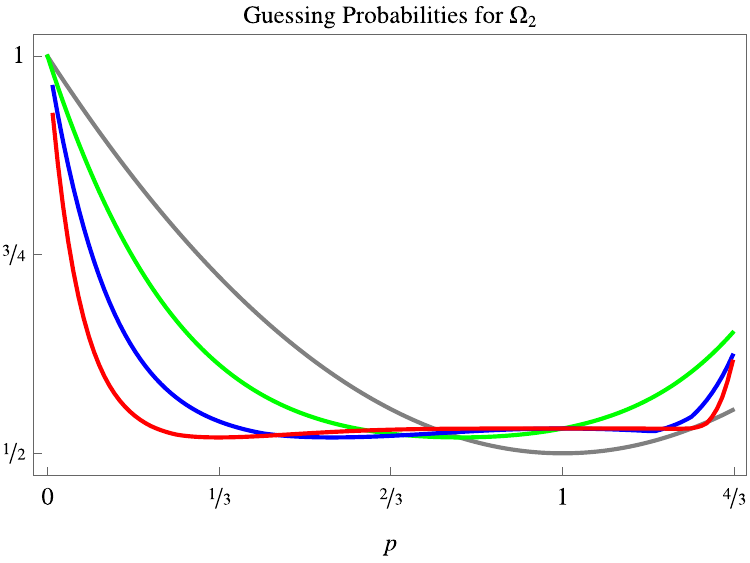} & \includegraphics[width=0.32\linewidth]{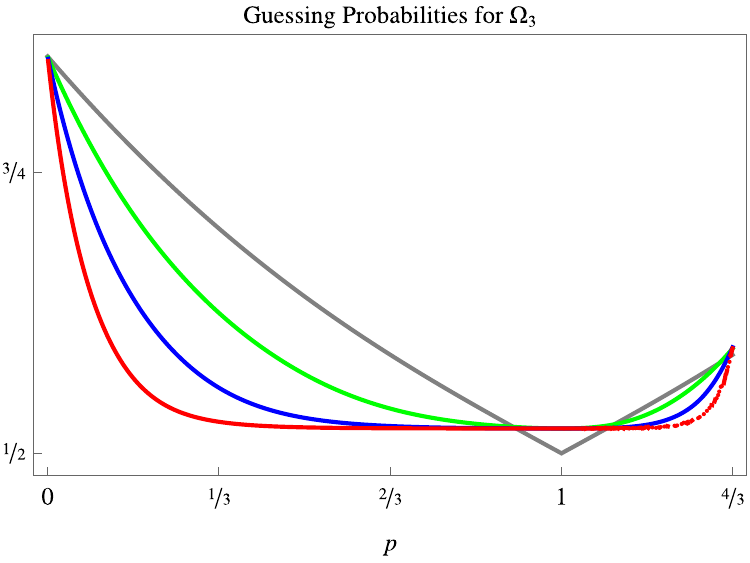}  \\[6pt]
		(d) & (e) & (f) \\ [6pt]
	\end{tabular}
	\caption{The guessing probabilities for the channel $\Delta_{p,q}(\rho)$ (grey), the quantum switch (green), the first-order superswitch (blue) and the second-order superswitch(red) for the three ensembles: (a) $\Omega_1$, (b) $\Omega_2$ and (c) $\Omega_3$. Sections of the 3D plots with $p=q$ are shown for the three ensembles: (d) $\Omega_1$, (e) $\Omega_2$ and (f) $\Omega_3$.}
	\label{fig: dim 4 depol var 3d}
\end{figure}

\begin{figure}[!t]
	\begin{tabular}{cc}
		\includegraphics[width=0.38\linewidth]{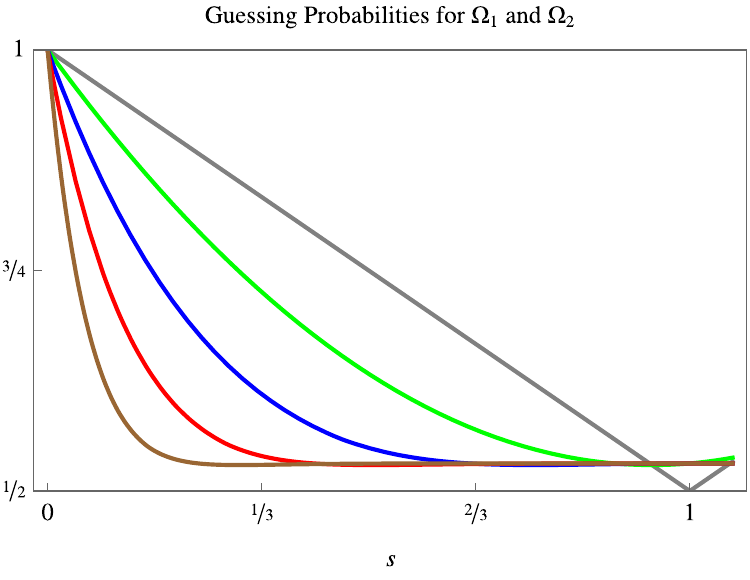} &    \includegraphics[width=0.555\linewidth]{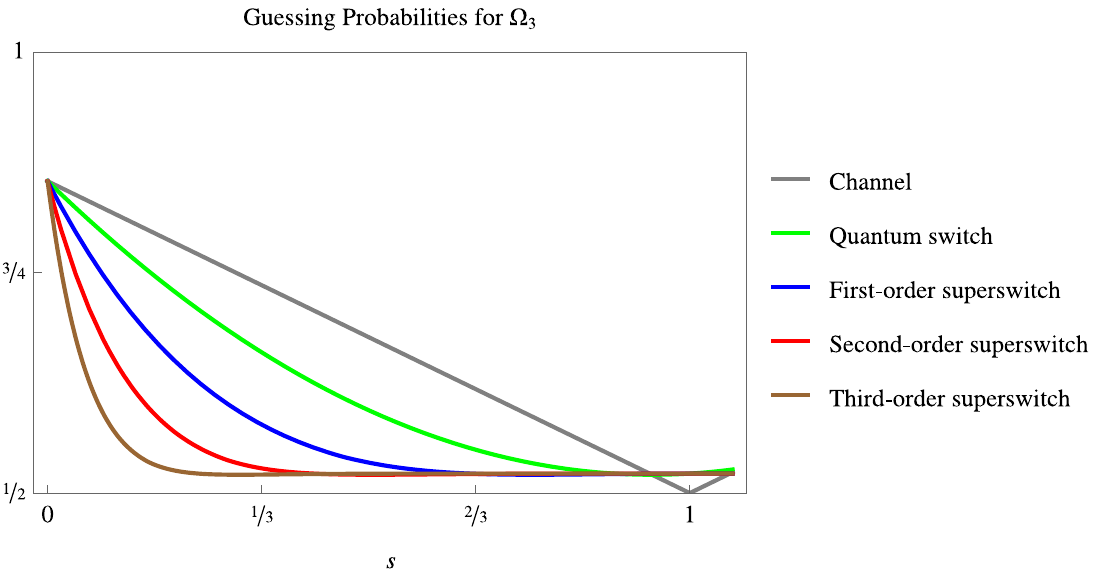}  \\[6pt]
		(a) & (b) \\ [6pt]
	\end{tabular}
	\caption{Plots of the guessing probabilities of the depolarisation channel $D_s(\rho)$ and all superswitches of order up to three. The guessing probabilities for ensembles $\Omega_1$ and $\Omega_2$ coincide and are shown jointly in (a), while for ensemble $\Omega_3$ in (b).}
	\label{fig: dim 4 depol}
\end{figure}

It follows that the two definitions coincide, i.e.~$\Delta_{p,q} = \D_{s}$, for the values $p=q=s=0,1$, which correspond to the identity channel, $\id$, and the completely depolarising channel, $\Lambda$, respectively.

Fig.\@ \ref{fig: dim 4 depol var 3d} shows the guessing probabilities for the channel $\Delta_{p,q}(\rho)$,  the quantum switch and first-order superswitch for ensembles (a) $\Omega_{1}=\left\{\nicefrac{1}{2}, \ketbra{xx}\right\}_{x=0,1}$, consisting of a pair of orthogonal product states, (b) $\Omega_{2}=\left\{\nicefrac{1}{2}, \ketbra{\Phi_i}\right\}_{i=\pm}$, with $\ket{\Phi_\pm} = \nicefrac{(\ket{00} \pm\ket{11})}{\sqrt{2}}$, consisting of two Bell states, and (c) $\Omega_{3}=\left\{\nicefrac{1}{2}, \ketbra{\psi_i}\right\}_{i=1,2}$, with $\ket{\psi_1} =\ket{00}$ and $\ket{\psi_1} =\ket{\Phi_-}$, consisting of two non-orthogonal states. In Fig.\@ \ref{fig: dim 4 depol var 3d}, we also show graphs of the guessing probabilities for $p=q$.

We repeat the analysis for the depolarisation channel $\D_{p}$, and plot the guessing probabilities in Fig.\@ \ref{fig: dim 4 depol}. The graphs follow a similar pattern to those of the channel $\Delta_{p,q}$ but there are some differences in the range where an improvement by the quantum switch or higher-order superswitches occurs. We find that the advantages offered by the superswitches are marginal and only for a small range of the depolarisation parameter $s$. Interestingly, for both ensembles $\Omega_1$ and $\Omega_2$ the guessing probabilities coincide, in contrast to the case of the channel $\Delta_{p,q}$ for $p=q$ where the region of improvement by the superswitches for the ensemble $\Omega_2$ is significantly smaller than that of $\Omega_1$. 

\begin{figure}[!b]
	\begin{tabular}{ccc}
		\includegraphics[width=0.32\linewidth]{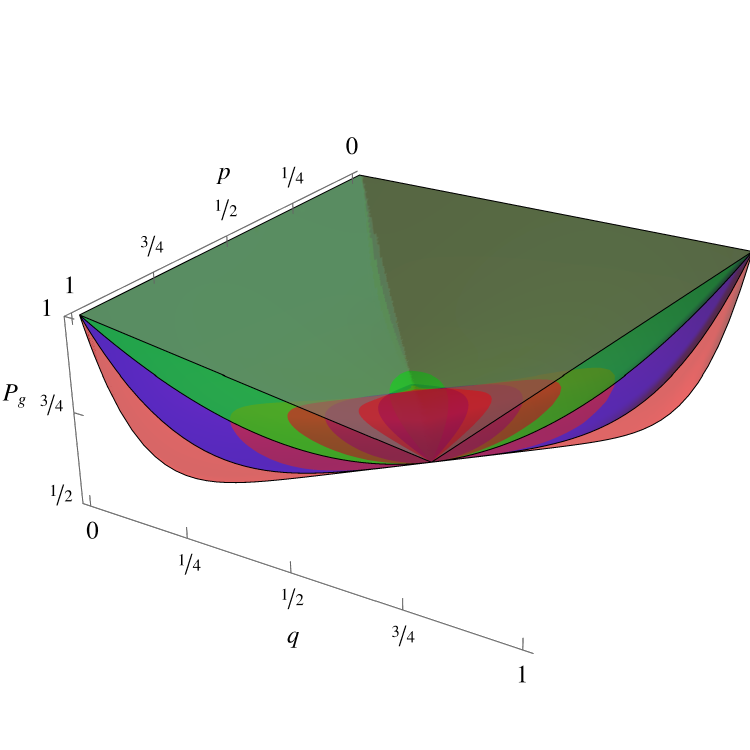} &   \includegraphics[width=0.32\linewidth]{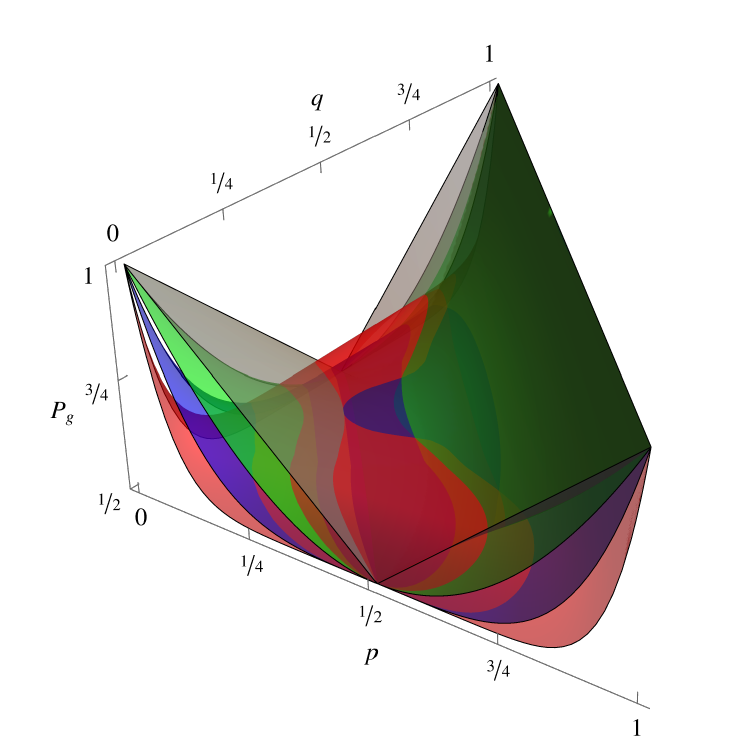} & \includegraphics[width=0.32\linewidth]{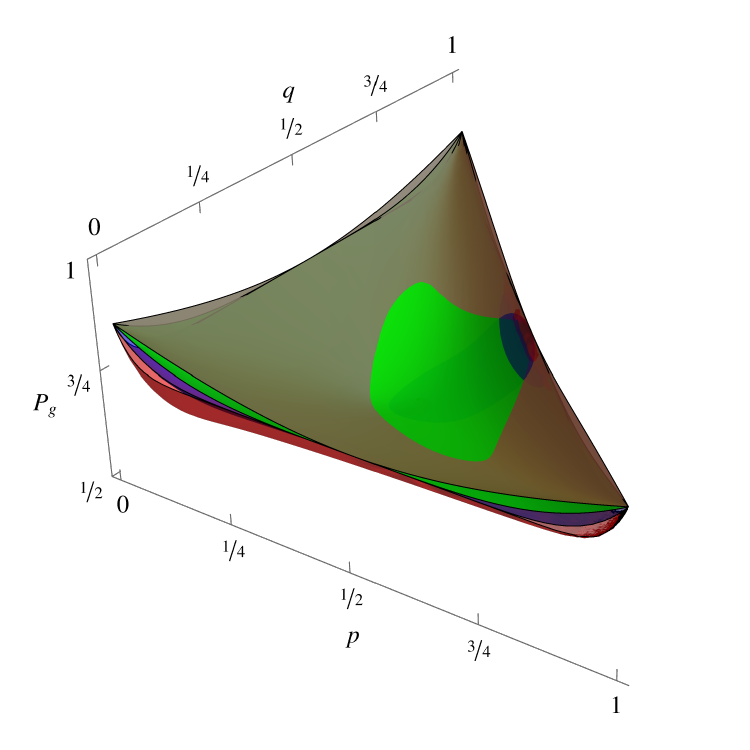}  \\[6pt]
		(a) & (b) & (c) \\ [6pt]
		\includegraphics[width=0.32\linewidth]{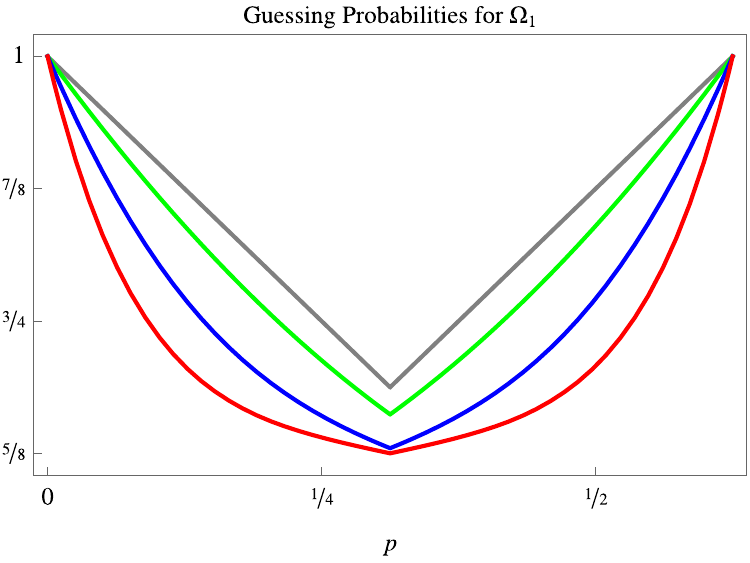} &   \includegraphics[width=0.32\linewidth]{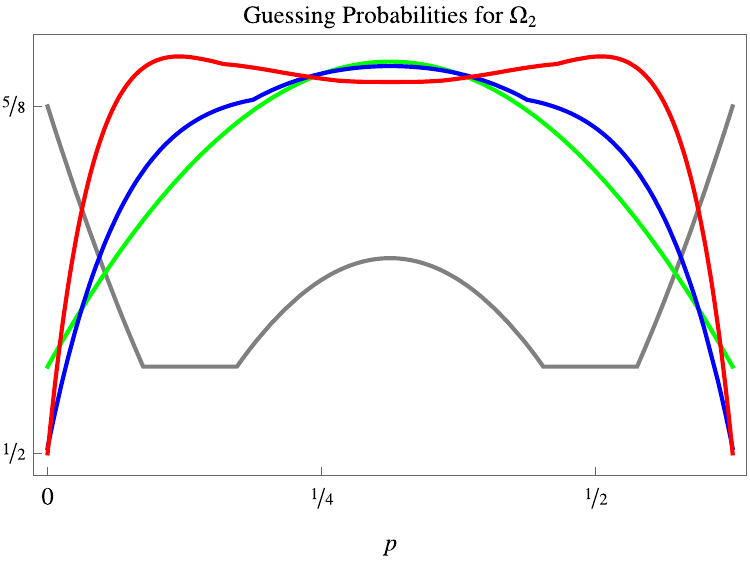} & \includegraphics[width=0.32\linewidth]{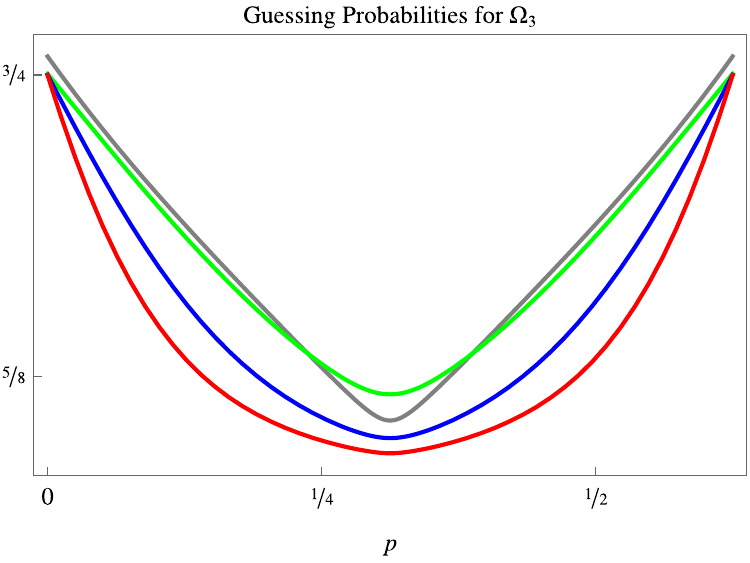}  \\[6pt]
		(d) & (e) & (f) \\ [6pt]
	\end{tabular}
		\caption{The guessing probabilities for the channel $\mathcal{W}_{p,q}(\rho)$ (grey), the quantum switch (green), the first-order superswitch (blue) and the second-order superswitch (red) for the three ensembles: (a) $\Omega_1$, (b) $\Omega_2$ and (c) $\Omega_3$. Sections of the 3D plots with $p+q=\nicefrac{5}{8}$ are shown for the three ensembles: (d) $\Omega_1$, (e) $\Omega_2$ and (f) $\Omega_3$.}
	\label{fig: dim 4 3d W}
\end{figure}

Finally, we consider the channel $\mathcal{W}_{q,p}(\rho)=\sum_i \vec{w}_i R_i \rho R_i ^\dagger$ defined by the probability vector 
\begin{align}
\vec{w}=\left\{(1-p-q)^2,0,(1-p-q)q,(1-p-q)p,0,0,0,0,p(1-p-q),0,pq,p^2,q(1-p-q),0,q^2,pq\right\} \,.
\end{align}
The action of the channel on the first part of the state is as follows: a $Y$ is applied with probability $p$, a $Z$ with probability $q$, or the state is left unchanged with probability $1-p-q$. On the second part of the state the probabilities are reversed, i.e.~the channel applies a $Y$ with probability $q$, a $Z$ with probability $p$, or leaves the state unchanged with probability $1-p-q$. 
In Fig.\@ \ref{fig: dim 4 3d W}, we show 3D plots of the guessing probabilities of the three ensembles $\Omega_1 ,\Omega_2$ and $\Omega_3$. For the two orthogonal product states, $\Omega_1$, there is a very small region of improvement for the quantum switch and a somewhat larger improvement for the second-order superswitch. For the pair of Bell states, $\Omega_2$, however, the quantum switch and the first- and second-order superswitches dominate the majority of the parameter space. Finally, in the case of the non-orthogonal pair, $\Omega_3$, only the quantum switch offers some improvement for small regions of the parameter space.
In Fig.\@ \ref{fig: dim 4 3d W}, we also show sections of the 3D plots for $p,q$ obeying $p+q=\nicefrac{5}{8}$. For the ensemble $\Omega_1$, there is no advantage in using any of the superswitches, while for ensemble $\Omega_3$ there is a small region where the quantum switch achieves a higher guessing probability. Ensemble $\Omega_3$ exhibits completely different behaviour, with the superswitches achieving a higher guessing probability in most of the region.


\section{Conclusion}
Quantum state discrimination can be formulated as a communication scenario between two parties over a quantum channel. In the traditional formulation, the channel is taken to be the identity.
We have studied a modification of this scenario, allowing for a possibly unknown channel to act between the two parties. Instead of performing the standard discrimination measurement, the decoding party employs a protocol based on a supermap, i.e.~the quantum switch, implemented by a communication provider that aids the standard quantum communication line. After a state is sent through the quantum switch, an appropriate discrimination measurement is applied, leading to the guessing of the label of the received states. We have found that employing such a protocol in many cases has two advantages: (i) an enhancement of the guessing probability, and (ii) knowledge of what optimal measurement to apply without assuming knowledge of the channel. 
In the first case, we have found that the guessing probability can increase even when comparing to multiple-copy state discrimination. In particular, the single-copy discrimination protocol with the quantum switch for near completely depolarising channels, achieves higher guessing probability than multiple-copy discrimination for any finite number of copies. The second advantage saves the cost of having to perform state or channel tomography in order to know what optimal measurement to apply, both of which are costly. We have also shown that, counter-intuitively, discrimination with a guessing probability $\pg=\nicefrac{5}{8}>\nicefrac{1}{2}$ can be performed for a pair of orthogonal states that are sent through a channel that is completely depolarising, i.e.~it maps every state to the maximally mixed one.

We have also defined higher-order switches, or superswitches, which when used for a communication scenario over a noisy channel can lead to a further increase in the guessing probability. We have derived a recurrence relation which can be used to iteratively calculate a superswitch up to any desired order.
We have explicitly evaluated superswitches of up to order four for various channels and found that their behaviour in terms of guessing probability is not trivial, that is, it is not always the case that a higher-order superswitch will outperform lower-order ones. In fact, for certain Pauli channels studied in this work, only some regions of the parameter space follow such a pattern. In the rest of the parameter space, an intricate picture emerges.
For a pair of orthogonal states, we have considered the set of all Pauli channels and found that: (i) for 34.6\% of the channels a protocol with a superswitch of order less than or equal to two cannot increase the guessing probability, (ii) for 45\% of them the quantum switch achieves the highest guessing probability, (iii) for 0.3\% of them the first-order superswitch achieves the highest guessing probability, and (iv) for 20.1\% of them the second-order superswitch achieves the highest guessing probability. 

Finally, we have defined superswitches for Pauli channels in any dimension $d=2^n$, with $n$ any positive integer and have explicitly studied superswitches of order up to two in dimension $d=4$ for a number of different channels. Similar to the case of dimension two, we found a non-trivial behaviour between the superswitches, with regions of parameter space where the channel achieves the highest guessing probabilities and other regions where a certain order superswitch dominates over the others.

A few open questions remain. Even though we found an improvement in the guessing for some ensembles and channels, we also demonstrated that there exist others for which the protocol makes the guessing worse. It would be useful to derive a more general characterisation in terms of mathematical conditions that detect whether the protocol with the superswitches can lead to an improvement or not, given some ensemble and channel.
In addition, even though we found that in many cases the higher-order switches offer an advantage in terms of the guessing probability in comparison to that of the channel or to the protocol with the quantum switch, for a significant range of the parameter space of the channels that we considered, the quantum switch still performs the best. It is worth asking whether there exist tasks beyond quantum state discrimination that are impossible with the quantum switch or other known supermaps, but are made possible with higher-order superswitches.

\section{Acknowledgments}
S.K., J.M.  and H.K. are supported by KIAS individual grant numbers CG086202 (S.K.), QP088701 (J.M.) and CG085301 (H.K.) at the Korea Institute for Advanced Study. AK and CL are supported by Institute for Information \& communications Technology Planning \& Evaluation (IITP) grant funded by the Korea government (MSIT) (RS-2023-00222863)

\bibliography{state_discrimination_with_indefinite_casaul_order_bib}
\bibliographystyle{apsrev4-2}

\appendix
\newpage

\section{Appendix A: Some results on minimum-error state discrimination with depolarisation noise}
Given an ensemble of states $\Omega = \{q_i, \rho_i\}_{i=1,\ldots,n}$, the guessing probability is found after maximising over all possible measurements, according to
\begin{align}
    \pg = \max_{\Pi} \sum_i q_i \Pr(i|\rho_i)= \max_{\Pi} \sum_i q_i \tr(\Pi_i \rho_i )
\end{align}
Note that an optimal POVM is in general not unique. For example, for the BB84 ensemble \begin{align}
	S^{BB84}=\left\{q_i=\nicefrac{1}{4},\{\proj{0},\proj{1},\proj{+},\proj{-}\}\right\}\,,
\end{align} 
a POVM of the form 
\begin{align}
    M=\left\{\alpha \proj{0}, \alpha \proj{1}, (1-\alpha)\proj{+},(1-\alpha)\proj{-}\right\}
\end{align}
achieves the optimal guessing probability of $\nicefrac{1}{2}$ for any value of $\alpha \in [0,1]$. For the values $\alpha=0$ and $\alpha=1$ only two states are ever identified by the measurement.
Thus, even though in the definition of the POVM we have in principle $n$ elements, some of them may be the null operator. In fact, in dimension two it is known that the optimal guessing probability can always be achieved with a POVM that identifies at most four states in the ensemble.

We now restrict to the case of ensembles with equal \emph{a priori} probabilities $q_i=\nicefrac{1}{n}$, $\Omega=\left\{\nicefrac{1}{n},\rho_i\right\}_{i=1}^n$ and let 
$\Pi=\{\Pi_i\}_{i=1}^n$ denote a POVM that corresponds to an optimal measurement. Restricting to dimension two and expressing everything in terms of Bloch vectors, the states are $\rho_i=\frac{1}{2}(\Id+\vec{r}_i \cdot \vec{\sigma})$ and the POVM elements are given by $\Pi_i = \sum_i \frac{w_i}{2}(\Id + \vec{m}_i\cdot \vec{\sigma})$. Moreover, since the operators $\Pi_i$ form a POVM we have that $\sum_i \Pi_i = \Id$ from which we obtain the conditions
\begin{align}
     \sum_i w_i =2 \,\, \text{and} \,\, \sum_i w_i \vec{m}_i = 0\,.
\end{align}
The guessing probability becomes 
\begin{align}
    \pg = \frac{\sum_i w_i \vec{m}_i \cdot \vec{r}_i}{2n} + \frac{1}{n} \,,
\end{align}
from which we obtain
\begin{align}
    \sum_i w_i \vec{m}_i \cdot \vec{r}_i  = 2(n\pg - 1) \,. \label{eq:sum to gp}
\end{align}

Let us consider the case where the states are sent through an unknown depolarisation channel as in Eq.\@ \eqref{eq:depolarisation}, producing the new ensemble $\Omega^{(\D)}=\left\{\nicefrac{1}{n},\D(\rho_i)\right\}_{i=1}^n$. It is known that the depolarisation channel is optimal measurement preserving for values of the parameter $p\in [0,1)$ \cite{kechrimparis_preserving_2018,kechrimparis_channel_2019}. That is, an optimal measurement of the original ensemble $\Omega$ is optimal for the new ensemble $\Omega^{(\D)}$ as well. On the other hand, if $p\in (1,\nicefrac{4}{3}]$, the optimal measurement is obtained from that of $\Omega$ by flipping the Bloch vector of each operator $\Pi_i$ in the POVM \cite{kechrimparis_channel_2019}. Explicitly,
\begin{align}
    \Pi_i\rightarrow \tilde{\Pi}_i = \frac{w_i}{2}\left(\Id-\vec{m}_i\cdot\vec{\sigma}\right)
\end{align}
In the case of two states, for example, this amounts to just relabeling the identified states as $1\leftrightarrow 2$. In other cases with similar symmetries, as in the BB84 or 6-state ensembles, it is again just a simple relabeling of outcomes but this is not the case in general. Nevertheless, in all cases the optimal measurement can be prepared if we know in advance the parameter range that $p$ takes values in. However, if we do not have any information on the value of $p$ applying the optimal measurement of the original ensemble will lead to a guessing that is in general worse than uniform. As a result, one would need to have an estimate of the parameter in the depolarisation channel in order to know which measurement, $\Pi$ or $\tilde{\Pi}$, to apply.
If we restrict to the case with $p\in[0,1]$, the guessing probability for the ensemble $\Omega^{(\D)}$ becomes
\begin{align}
    \pg^{(\D)} &= \frac{1}{n}\sum_i \tr(\Pi_i \D(\rho_i)) \notag \\
    &= \frac{1}{n} \sum_i \tr\left(\frac{1}{2}(\Id+(1-p)\vec{r}_i \cdot \vec{\sigma})\frac{w_i}{2}(\Id + \vec{m}_i\cdot \vec{\sigma})\right) \notag \\
    &= \frac{1}{n} \sum_i \frac{w_i}{4} \tr\left((\Id+(1-p)\vec{r}_i \cdot \vec{\sigma}) (\Id + \vec{m}_i\cdot \vec{\sigma})\right) \notag \\
    &= \frac{1}{n} \sum_i \frac{w_i}{4} \left(2+2(1-p)\vec{r}_i\cdot \vec{m}_i\right) \notag \\
    &= \frac{1}{n} + \frac{(1-p)}{2n}\sum_i w_i \vec{r}_i\cdot \vec{m}_i \notag \\
    &= \frac{1}{n} + \frac{(1-p)}{n} (n\pg - 1) \,,
\end{align}
from which we obtain 
\begin{align}
    \pg^{(\D)} = (1-p) \pg +\frac{p}{n}\, \,, \,\, p\in[0,1]\,. \label{eq:guessing prob depolarisation}
\end{align}
In the case of $p\in (1,\nicefrac{4}{3}]$, we have to use the optimal measurement $\tilde{\Pi}$, which changes some signs, and the expression becomes
\begin{align}
    \pg^{(\D)} = (p-1) \pg +\frac{2-p}{n}\, \,, \,\,  p\in(1,\nicefrac{4}{3}] \,. \label{eq:guessing prob depolarisation flipped}
\end{align}

\section{Appendix B: Helstrom bound for qubit states}
We rewrite the Helstrom bound \cite{helstrom_quantum_1969} for a pair of qubit states in terms of their Bloch vectors. Specifically, given an ensemble of two qubit states $\Omega=\{q_i, \rho_i\}_{i=1,2}$, with $\rho_i=(\Id+\vec{r}_i \cdot \vec{\sigma})/2$ where $\vec{r}_i$ are their Bloch vectors, the Helstrom bound for the optimal guessing probability is
\begin{align}
	\pg =  \frac{1}{2} +\frac{\norm{h}_1}{2}= \frac{1}{2} +\frac{\norm{q_1 \rho_1-q_2 \rho_2}_1}{2},
\end{align}
where we have defined the Helstrom operator $h=q_1 \rho_1-q_2 \rho_2$ and with $\norm{h}_1=\tr \sqrt{h h^\dagger}$ denoting the trace norm. Explicit evaluation of the trace norm gives
\begin{align}
	\norm{h}_1 &= \frac{1}{2}\abs{q_1-q_2} +\frac{1}{2}\norm{q_1 \vec{r}_1-q_2 \vec{r}_2}_2+\frac{1}{2}\abs{\abs{q_1-q_2}-\norm{q_1 \vec{r}_1-q_2 \vec{r}_2}_2} \notag \\
	&= \max\left\{\abs{q_1-q_2}, \norm{q_1 \vec{r}_1-q_2 \vec{r}_2}_2 \right\} \,.
\end{align}
We note that the case $\abs{q_1-q_2}\geq \norm{q_1 \vec{r}_1-q_2 \vec{r}_2}_2$ corresponds to the optimal guessing probability being achieved without performing a measurement and always guessing the state with the the highest \emph{a priori} probability. In the case of equal probabilities, $q_1=q_2=\nicefrac{1}{2}$, we obtain
\begin{align}
	\norm{h}_1 = \frac{\norm{ \vec{r}_1- \vec{r}_2}_2 }{2}\,,
\end{align}
and in this case the guessing probability becomes
\begin{align}
	\pg =  \frac{1}{2} +\frac{\norm{ \vec{r}_1- \vec{r}_2}_2 }{4} \,.
\end{align}

\section{Appendix C: Helstrom bound for multiple copies of orthogonal states under depolarisation noise}
We prove the general expression for the Helstrom bound for multiple copies of a pair of orthogonal states under depolarisation noise. Any pair of orthogonal states is unitarily equivalent to the pair $\ketbra{0}, \ketbra{1}$. If these states are sent through a depolarisation channel $\D_p(\rho)$, their Bloch vector is transformed as $\vec{r}_i\rightarrow (1-p)\vec{r}_i$, or 
\begin{align}
 \D_p(\ketbra{0}) = \begin{pmatrix}
		1-\nicefrac{p}{2} & 0 \\
		0 &\nicefrac{p}{2}
	\end{pmatrix} \,, \,\,\,
 	\D_p(\ketbra{1}) = \begin{pmatrix}
		\nicefrac{p}{2} & 0 \\
		0 &1-\nicefrac{p}{2}
	\end{pmatrix} \,.
\end{align}
The Hestrom bound for $n$ copies of these states appearing with equal \emph{a priori} probabilities is
 \begin{align}
	\pg^{(\D_p ,n)} &= \frac{1}{2}+\frac{1}{4}\norm{\D_p(\rho_0)\otimes \underset{n}{\cdots} \otimes\D_p(\rho_0)-\D_p(\rho_1)\otimes \underset{n}{\cdots} \otimes\D_p(\rho_1)}_1 \,.
\end{align}
To evaluate the expression, we first consider the matrix
\begin{align}
	M_0 (a) = \begin{pmatrix}
		a & 0 \\
		0 & 1-a
	\end{pmatrix} \,.
\end{align}
It is clear that since both matrices are diagonal, their Kronecker products are also diagonal. Moreover, given that the trace norm of a diagonal matrix with real entries is the sum of absolute values of the diagonal entries , it remains to correctly list all terms in the diagonal for the general case of $n$ Kronecker products. We have,
\begin{align}
	M_n(a) = M(a) \underset{n}{\otimes \ldots \otimes} M(a) = M(a)\otimes M_{n-1}(a)= \begin{pmatrix}
		a M_{n-1}(a) & \mathbf{0}_{n-1} \\
		 \mathbf{0}_{n-1} & (1-a) M_{n-1}(a) \,.
	\end{pmatrix}
\end{align}
where $ \mathbf{0}_{n}$ denotes a $2^n \times 2^n$ matrix with all entries 0.
We note that all terms of order $n$ are generated in the same way as all the numbers that can be represented in binary with $n$ digits, that is, by appending 0's and 1's to the previous orders, up to a rearrangement of terms. For example, for $n=3$, we can represent the numbers $0,1,2,3,4,5,6,7$ in binary as $000,001,010,011,100,101,111$. If we let $0\rightarrow a$ and $1\rightarrow 1-a$ and consider the multiplication of terms, as in for instance $010 = a(1-a)a = a^2 (1-a)$, we get all terms in the diagonal but rearranged. 
The ordering is not important, however, since when we take the trace norm we take the sum of the absolute values of all terms. Finally, by noting that
\begin{align}
	\D_p(\rho_0)\otimes \underset{n}{\cdots} \otimes\D_p(\rho_0) = M_n(1-\nicefrac{p}{2})\,,
\end{align} 
and 
\begin{align}
	\D_p(\rho_1)\otimes \underset{n}{\cdots} \otimes\D_p(\rho_1) = M_n(\nicefrac{p}{2})\,,
\end{align} 
which implies that each term in $\D_p(\rho_1)\otimes \underset{n}{\cdots} \otimes\D_p(\rho_1)$ is the same as in $\D_p(\rho_0)\otimes \underset{n}{\cdots} \otimes\D_p(\rho_0)$ after the transformation $\nicefrac{p}{2}\rightarrow 1-\nicefrac{p}{2}$, we finally obtain the expression
 \begin{align}
	\pg^{(\D_p ,n)} = \frac{1}{2}+\frac{1}{4} \sum_{k=0}^{n} \frac{n!}{k! (n-k)!}\abs{\left(\frac{p}{2}\right)^k \left(1-\frac{p}{2}\right)^{n-k}-\left(\frac{p}{2}\right)^{n-k} \left(1-\frac{p}{2}\right)^k}\,. \label{eq: Helstrom multiple orth depol}
\end{align}

\section{Appendix D: Entangled states in the ancillas}
We show that by letting the ancillas of the innermost switches be entangled, the general form of the first-order superswitch reproduces the one in \cite{das_quantum_2022}. First we rewrite the Kraus operators of a superswitch where there is only one control for both of the inside switches and we denote them by $\tilde{K}_{ijkl}$. Specifically,
\begin{align}
	\tilde{K}_{ijkl}=  &E_i F_j \ee_k \ff_l \otimes \ketbra{0}\otimes \ketbra{0} +  F_j E_i \ff_l  \ee_k  \otimes \ketbra{1}\otimes \ketbra{0} \notag \\
	+&\ee_k \ff_l  E_i F_j\otimes \ketbra{0}\otimes \ketbra{1} + \ff_l  \ee_k  F_j E_i   \otimes \ketbra{1}\otimes \ketbra{1} \notag \\
	=&\frac{1}{8}\left( \acom{\acom{E_i}{F_j}}{\acom{\ee_k}{\ff_l}} +  \acom{\com{E_i}{F_j}}{\com{\ee_k}{\ff_l}}  \right) \otimes \Id \otimes \Id 
	\notag \\
	+&\frac{1}{8}\left(\acom{\acom{E_i}{F_j}}{\com{\ee_k}{\ff_l}}  + \acom{\com{E_i}{F_j}}{\acom{\ee_k}{\ff_l}}  \right)\otimes Z \otimes \Id \notag \\
	+&\frac{1}{8}\left( \com{\acom{E_i}{F_j}}{\acom{\ee_k}{\ff_l}}  + \,\, \com{\com{E_i}{F_j}}{\com{\ee_k}{\ff_l}}  \right)\,\,\otimes \Id \otimes Z
	\notag \\
	+&\frac{1}{8}\left(\com{\acom{E_i}{F_j}}{\com{\ee_k}{\ff_l}} \,\, \,+ \,\com{\com{E_i}{F_j}}{\acom{\ee_k}{\ff_l}}  \right)\otimes Z \otimes Z\,.
\end{align}
It follows that the correlated higher-order switch of \cite{das_quantum_2022}, which we denote by $\tilde{\Sw}^{(1)}$, becomes
\begin{align}
	\tilde{\Sw}^{(1)}_{\omega}&(\E,\F,\EE,\FF)(\rho)= \frac{1}{64}\sum_{i,j,k,l} \bigg[ \notag \\
	&\left(\acom{\acom{E_i}{F_j}}{\acom{\ee_k}{\ff_l}}\rho \acom{\acom{E_i}{F_j}}{\acom{\ee_k}{\ff_l}}^\dagger +\acom{\com{E_i}{F_j}}{\com{\ee_k}{\ff_l}}\rho \acom{\com{E_i}{F_j}}{\com{\ee_k}{\ff_l}}^\dagger \right)\otimes  \omega_c  \notag \\
	+&\left(\acom{\acom{E_i}{F_j}}{\com{\ee_k}{\ff_l}}\rho \acom{\acom{E_i}{F_j}}{\com{\ee_k}{\ff_l}}^\dagger +\acom{\com{E_i}{F_j}}{\acom{\ee_k}{\ff_l}}\rho \acom{\com{E_i}{F_j}}{\acom{\ee_k}{\ff_l}}^\dagger \right)\otimes (Z \otimes \Id)  \omega_c (Z \otimes \Id)  \notag \\
	+&\left(\com{\acom{E_i}{F_j}}{\acom{\ee_k}{\ff_l}}\rho \com{\acom{E_i}{F_j}}{\acom{\ee_k}{\ff_l}}^\dagger +\com{\com{E_i}{F_j}}{\com{\ee_k}{\ff_l}}\rho \com{\com{E_i}{F_j}}{\com{\ee_k}{\ff_l}}^\dagger \right)\otimes (\Id \otimes Z)  \omega_c (\Id \otimes Z)  \notag \\
	+&\left(\com{\acom{E_i}{F_j}}{\com{\ee_k}{\ff_l}}\rho \com{\acom{E_i}{F_j}}{\com{\ee_k}{\ff_l}}^\dagger +\com{\com{E_i}{F_j}}{\acom{\ee_k}{\ff_l}}\rho \com{\com{E_i}{F_j}}{\acom{\ee_k}{\ff_l}}^\dagger \right)\otimes (Z \otimes Z)  \omega_c (Z \otimes Z)  \bigg]\,, \notag \\		
\end{align}
and if we choose the state of the controls to be $\omega_c= \ketbra{+} \otimes \ketbra{+} $, we obtain 
\begin{align}
	\tilde{\Sw}^{(1)}_{\omega}&(\E,\F,\EE,\FF)(\rho)= \frac{1}{64}\sum_{i,j,k,l} \bigg[ \notag \\
	&\left(\acom{\acom{E_i}{F_j}}{\acom{\ee_k}{\ff_l}}\rho \acom{\acom{E_i}{F_j}}{\acom{\ee_k}{\ff_l}}^\dagger +\acom{\com{E_i}{F_j}}{\com{\ee_k}{\ff_l}}\rho \acom{\com{E_i}{F_j}}{\com{\ee_k}{\ff_l}}^\dagger \right)\otimes \ketbra{++}  \notag \\
	+&\left(\acom{\acom{E_i}{F_j}}{\com{\ee_k}{\ff_l}}\rho \acom{\acom{E_i}{F_j}}{\com{\ee_k}{\ff_l}}^\dagger +\acom{\com{E_i}{F_j}}{\acom{\ee_k}{\ff_l}}\rho \acom{\com{E_i}{F_j}}{\acom{\ee_k}{\ff_l}}^\dagger \right)\otimes \ketbra{-+}   \notag \\
	+&\left(\com{\acom{E_i}{F_j}}{\acom{\ee_k}{\ff_l}}\rho \com{\acom{E_i}{F_j}}{\acom{\ee_k}{\ff_l}}^\dagger +\com{\com{E_i}{F_j}}{\com{\ee_k}{\ff_l}}\rho \com{\com{E_i}{F_j}}{\com{\ee_k}{\ff_l}}^\dagger \right)\otimes \ketbra{+-}   \notag \\
	+&\left(\com{\acom{E_i}{F_j}}{\com{\ee_k}{\ff_l}}\rho \com{\acom{E_i}{F_j}}{\com{\ee_k}{\ff_l}}^\dagger +\com{\com{E_i}{F_j}}{\acom{\ee_k}{\ff_l}}\rho \com{\com{E_i}{F_j}}{\acom{\ee_k}{\ff_l}}^\dagger \right)\otimes \ketbra{--}   \bigg]\,, \notag \\		\label{eq: 1st order superswitch correlated}
\end{align}
and the four channels can be fully separated by performing measurements in the $\ketbra{\pm}$ basis.

At the same time, it is not hard to see that by letting $\omega_c=\omega_{12} \otimes \omega$ with $\omega_{12} = \ketbra{\Phi^+}$ and $\omega=\ketbra{+}$ in Eq.\@ \eqref{eq: 1st order superswitch} and using the relations
\begin{align}
	(\Id \otimes Z)  \ket{\Phi^+} = (\Id \otimes Z)  \ket{\Phi^+} = \frac{1}{\sqrt{2}} \left(\ket{00}-\ket{11}\right) = \ket{\Phi^-}\,, \, \, \text{and} \,\, (Z \otimes Z) \ket{\Phi^+} =  \ket{\Phi^+}  \,,
\end{align} 
we obtain the same expression as in Eq.\@ \eqref{eq: 1st order superswitch correlated} but with the four ancilla states being $\ketbra{\Phi^+ +}\equiv \ketbra{\phi^+}\otimes \ketbra{+}$, $\ketbra{\Phi^- +},$ $\ketbra{\Phi^+ -},\ketbra{\Phi^- -}$ in place of $\ketbra{++}, \ketbra{-+}, \ketbra{+-}, \ketbra{--}$, respectively. Then, be performing a bell measurement on the ancillas that control the inside switches and a measurement in the $\ketbra{\pm}$ basis achieves the same result as the correlated switch in Eq.\@ \eqref{eq: 1st order superswitch correlated}.

\section{Appendix E: Derivation of the recurrence relation and update rule}
The general expression for the superswitches, Eq.\@ \eqref{eq: general switch}, involves terms with nested commutators and anticommutators of Kraus operators of the channels. Given the restriction to Pauli channels and the fact that commutators and anticommutators of the four Pauli matrices, $\Id, X, Y, Z$, are also Pauli matrices allows to derive a recurrence relation and an update rule that makes possible to evaluate the next order superswitch given the previous one. The reason for that is that at a given order after measurements on the ancilla qubits, the channel that acts on the state is necessarily a Pauli channel. In the next order, if we consider all pairs of channels that result from the previous order and evaluate their commutator and anticommutators (controlled by the extra ancilla qubit in the next order), we can derive a general rule for the update from one order to the next. Essentially, since we have defined switch of switches, each order follows by evaluating the quantum switch of all possible channels generated in previous orders.

In detail, let $\E(\rho)=\sum_i \vec{r}_{1,i} \sigma_i \rho \sigma_i$ and $\F(\rho)=\sum_i \vec{r}_{2,i} \sigma_i \rho \sigma_i$, denote two Pauli channels with probability vectors $\vec{r}_i = \{\alpha_i, \beta_i, \gamma_i, \delta_i \}$. If we denote by $\vec{\sigma}=\{\Id,X,Y,X\}$ the `vector' of Pauli matrices in that explicit order and by gathering all anticommutators and commutators of Kraus operators in two matrices with operator-valued elements, $\mathcal{A}$ and $\mathcal{C}$, respectively, we find
\begin{align}
\mathcal{A}= 2\cdot
\begin{pmatrix}
	\sqrt{\alpha_1 \alpha_2} \, \Id & \sqrt{\alpha_1 \beta_2} \, X & \sqrt{\alpha_1 \gamma_2} \, Y & \sqrt{\alpha_1 \delta_2} \, Z \\
	\sqrt{\beta_1 \alpha_2} \, X & \sqrt{\beta_1 \beta_2} \, \Id &0 & 0 \\
	\sqrt{\gamma_1 \alpha_2} \, Y & 0 & \sqrt{\gamma_1 \gamma_2} \, \Id & 0 \\
	\sqrt{\delta_1 \alpha_2} \, Z &0 & 0 & \sqrt{\delta_1 \delta_2} \, \Id 
\end{pmatrix} \,,
\end{align}
and
\begin{align}
	\mathcal{C}= 2 \textrm{i} \cdot
	\begin{pmatrix}
		0 & 0 & 0 & 0 \\
	    0 & 0 & \,\,\,\,\,\sqrt{\beta_1 \gamma_2} \, Z & -\sqrt{\beta_1 \delta_2} \, Y \\
		0 & -\sqrt{\gamma_1 \beta_2} \, Z  & 0 & \,\,\,\,\,\,\sqrt{\gamma_1 \delta_2} \, X \\
		0 & \,\,\,\,\,\,\sqrt{\delta_1 \beta_2} \, Y & -\sqrt{ \delta_1 \gamma_2} \, X & 0 
	\end{pmatrix} \,.
\end{align}
Then, the anticommutator and commutator terms in the definition of the quantum switch term, Eq.\@ \eqref{eq: zero order ss- coms and anticoms}, can be concisely written 
\begin{align}
\frac{1}{4} \sum_{i,j} \left\{E_i, F_j \right\}\rho\left\{E_i, F_j \right\}^\dagger = e^\top \cdot \left( \mathcal{A} \circ \rho \circ \mathcal{A}^* \right) \cdot e 
\end{align}
and
\begin{align}
  \frac{1}{4} \sum_{i,j} \left[E_i, F_j\right]\rho \left[E_i, F_j\right] ^\dagger = e^\top \cdot \left( \mathcal{C} \circ \rho \circ \mathcal{C}^* \right) \cdot e 
\end{align}
where $e=(1,1,1,1)$ denotes a vector of ones and `$\cdot$' denotes usual matrix multiplication. We have also defined the operation `$\ldots \circ \ldots \circ \ldots$' that takes two $4\times$ operator-valued matrices on the left and right side and a density matrix in the middle and returns a $4\times 4$ operator-valued matrix, as well as the operation $^{*}$ which denotes taking the conjugate transpose of each element of the operator-valued matrix. In more detail, the first denotes element-wise multiplication of the two outside matrices while `sandwiching' the density matrix $\rho$ with the operators on the left and right. For example, the second element of the first row of $\mathcal{A} \circ \rho \circ \mathcal{A}$ would give the contribution $\left[\mathcal{A} \circ \rho \circ \mathcal{A}^*\right]_{12}= \alpha_1 \beta _2 X\rho X$. The operation $^*$ is not important in the case we are considering, as the conjugate transpose of all Pauli matrices is equal to themselves and each element of $\mathcal{A}, \mathcal{C}$ consists of some real number multiplying one of the Pauli matrices. 
Finally, by taking left and right products with the vector with unit entries $e$, we get all terms in the summation. Explicitly, we find
\begin{align}
	\frac{1}{4} \sum_{i,j} \left\{E_i, F_j \right\}\rho\left\{E_i, F_j \right\}^\dagger =& \left(\alpha_1 \alpha_2 +\beta_1\beta_2+\gamma_1 \gamma_2 +\delta_1 \delta_2\right) \rho + \left(\alpha_1 \beta_2+\beta_1 \alpha_2\right)X\rho X \notag \\
	&+ \left(\alpha_1 \gamma_2+\gamma_1 \alpha_2\right)Y\rho Y+ \left(\alpha_1 \delta_2+\delta_1 \alpha_2\right)Z\rho Z \,,
\end{align}
and
\begin{align}
	\frac{1}{4} \sum_{i,j} \left[E_i, F_j\right]\rho \left[E_i, F_j\right] ^\dagger  =  &\left( \beta_1 \gamma_2+\gamma_1 \beta_2\right)X\rho X + \left( \gamma_1 \delta_2+\delta_1 \gamma_2\right)Y\rho Y\notag \\
	&+ \left(\delta_1 \beta_2+\beta_1 \delta_2\right)Z\rho Z \,.
\end{align}
By noting that the expressions do not give channels but channels multiplied by probabilities, the update rule, Eq.\@ \eqref{eq: update rule}, follows.

\section{Appendix F: Guessing probabilities for superswitches and the depolarisation channel}
For the ensemble of orthogonal states $\Omega=\{\nicefrac{1}{2},\ketbra{i}\}_{i=0,1}$ that go through a depolarisation channel, Eq. \eqref{eq:depolarisation}, that transforms the ensemble $\Omega$ to $\Omega^{(\D)}=\{\nicefrac{1}{2},\D_p (\ketbra{i})\}_{i=0,1}$ we list the guessing probabilities if no superswitches are used, $\pg^{(\D)}$, as well as the protocols with superswitches up to order two, $\pg^{(n)}$, where the index $n$ denotes the order of the superswitch. For the depolarisation channel we have the optimal guessing
\begin{align}
\pg^{(\D)}=\frac{1}{2}(1+\abs{1-p})\,,
\end{align}
while the quantum switch leads to a guessing given by
\begin{align}
	\pg^{(0)}=1-p+\frac{5p^2}{8}\,.
\end{align}
The next two superswitches give
\begin{align}
	\pg^{(1)}=1-2p+\frac{29 p^2}{8}-\frac{23 p^3}{8}+\frac{55 p^4}{64}+\frac{1}{64} p^2 | 3 p (5 p-8)+8| \,,
\end{align}
and
\begin{align}
\pg^{(2)}=&1-4p+\frac{67 p^2}{4}-\frac{159 p^3}{4}+\frac{1897 p^4}{32}-\frac{457 p^5}{8}+\frac{1975 p^8}{1024}+\frac{4457 p^6}{128}-\frac{1573
	p^7}{128} \notag \\
&+\frac{p^2
		| 9 p (p (p (3 p (p (71 p-360)+760)-2560)+1600)-512)+512| }{2048}\notag \\
&+\frac{1}{128} p^2 | 3 p (3 p (p (19 p-64)+80)-128)+64|  
-\frac{1}{128}
p^3 | 3 p (3 p (p (19 p-64)+80)-128)+64| \notag \\
&+\frac{1}{512} p^4 | 3 p (3 p (p (19 p-64)+80)-128)+64| +\frac{3 p^6 | 6 p (5 p-8)+16| }{4096} \,.
\end{align}
We omit the formulas for the guessing probabilities for higher-order superswitches as the expressions are long.

\section{Appendix G: The update rule in dimension four}
Here we list the update rules for Pauli channels in dimension four. After a similar calculation to that found in Appendix E, we explicitly obtain 

\begin{align}
	\acm(\vec{r}, \vec{v}) = \Big\{\,\,\,&\sum_{i=0}^{15}p_{i} q_{i}, 
 \,\,\, p_1 q_0+ p_0 q_1+ p_5 q_4+ p_4 q_5+ p_9 q_8+
p_8 q_9+ p_{13} q_{12}+ p_{12} q_{13}, \notag \\
 &p_2 q_0+p_0 q_2+ p_6 q_4+ p_4q_6+ p_{10} q_8+ p_8 q_{10}+ p_{14} q_{12}+ p_{12} q_{14}, \notag \\
 &p_3 q_0+ p_0q_3+ p_7 q_4+p_4 q_7+ p_{11} q_8+ p_8 q_{11}+ p_{15} q_{12}+ p_{12}q_{15}, \notag \\
 &p_4 q_0+ p_5 q_1+ p_6 q_2+ p_7 q_3+ p_0 q_4+ p_1 q_5+ p_2 q_6+p_3 q_7, \notag \\
&p_5 q_0+ p_4 q_1+ p_1 q_4+ p_0 q_5+ p_{15} q_{10}+ p_{14}
q_{11}+ p_{11} q_{14}+ p_{10} q_{15},\notag \\
 &p_6 q_0+ p_4 q_2+ p_2 q_4+ p_0 q_6+
p_{15} q_9+ p_{13} q_{11}+ p_{11} q_{13}+ p_9 q_{15},\notag \\
 &p_7 q_0+ p_4 q_3+p_3 q_4+ p_0 q_7+ p_{14} q_9+ p_{13} q_{10}+ p_{10} q_{13}+ p_9 q_{14},\notag \\
&p_8 q_0+ p_9 q_1+ p_{10} q_2+ p_{11} q_3+ p_0 q_8+ p_1 q_9+ p_2 q_{10}+p_3 q_{11},\notag \\
 &p_9 q_0+ p_8 q_1+ p_{15} q_6+ p_{14} q_7+ p_1 q_8+ p_0 q_9+p_7 q_{14}+ p_6 q_{15},\notag \\
& p_{10} q_0+ p_8 q_2+ p_{15} q_5+ p_{13} q_7+ p_2q_8+p_0 q_{10}+ p_7 q_{13}+ p_5 q_{15},\notag \\
 &p_{11} q_0+ p_8 q_3+ p_{14} q_5+
p_{13} q_6+ p_3 q_8+ p_0 q_{11}+ p_6 q_{13}+ p_5 q_{14},\notag \\
 &p_{12} q_0+p_{13} q_1+ p_{14} q_2+ p_{15} q_3+ p_0 q_{12}+ p_1 q_{13}+ p_2 q_{14}+p_3 q_{15},\notag \\
 &p_{13} q_0+ p_{12} q_1+ p_{11} q_6+ p_{10} q_7+ p_7 q_{10}+p_6 q_{11}+ p_1 q_{12}+ p_0 q_{13},\notag \\
 &p_{14} q_0+ p_{12} q_2+ p_{11} q_5+p_9 q_7+ p_7 q_9+ p_5 q_{11}+ p_2 q_{12}+ p_0 q_{14},\notag \\
&p_{15} q_0+ p_{12}q_3+ p_{10} q_5+ p_9 q_6+ p_6 q_9+ p_5 q_{10}+ p_3 q_{12}+ p_0q_{15} \Big\}
\end{align}
and
\begin{align}
	\cm(\vec{r}, \vec{v}) = \Big\{\,\,\, &0, \,\,\,\, p_3 q_2+p_2 q_3+p_7 q_6+p_6 q_7+p_{11} q_{10}+p_{10} q_{11}+p_{15}
q_{14}+p_{14} q_{15},\notag \\
&p_3 q_1+p_1 q_3+p_7 q_5+p_5 q_7+p_{11} q_9+p_9
q_{11}+p_{15} q_{13}+p_{13} q_{15},\notag \\
&p_2 q_1+p_1 q_2+p_6 q_5+p_5 q_6+p_{10}
q_9+p_9 q_{10}+p_{14} q_{13}+p_{13} q_{14},\notag \\
&p_{12} q_8+p_{13} q_9+p_{14}
q_{10}+p_{15} q_{11}+p_8 q_{12}+p_9 q_{13}+p_{10} q_{14}+p_{11} q_{15},\notag \\
&p_7
q_2+p_6 q_3+p_3 q_6+p_2 q_7+p_{13} q_8+p_{12} q_9+p_9 q_{12}+p_8 q_{13},\notag \\
&p_7
q_1+p_5 q_3+p_3 q_5+p_1 q_7+p_{14} q_8+p_{12} q_{10}+p_{10} q_{12}+p_8
q_{14},\notag \\
&p_6 q_1+p_5 q_2+p_2 q_5+p_1 q_6+p_{15} q_8+p_{12} q_{11}+p_{11}
q_{12}+p_8 q_{15},\notag \\
&p_{12} q_4+p_{13} q_5+p_{14} q_6+p_{15} q_7+p_4 q_{12}+p_5
q_{13}+p_6 q_{14}+p_7 q_{15},\notag \\
&p_{11} q_2+p_{10} q_3+p_{13} q_4+p_{12} q_5+p_3
q_{10}+p_2 q_{11}+p_5 q_{12}+p_4 q_{13},\notag \\
&p_{11} q_1+p_9 q_3+p_{14} q_4+p_{12}
q_6+p_3 q_9+p_1 q_{11}+p_6 q_{12}+p_4 q_{14},\notag \\
&p_{10} q_1+p_9 q_2+p_{15}
q_4+p_{12} q_7+p_2 q_9+p_1 q_{10}+p_7 q_{12}+p_4 q_{15},\notag \\
&p_8 q_4+p_9 q_5+p_{10}
q_6+p_{11} q_7+p_4 q_8+p_5 q_9+p_6 q_{10}+p_7 q_{11},\notag \\
&p_{15} q_2+p_{14} q_3+p_9
q_4+p_8 q_5+p_5 q_8+p_4 q_9+p_3 q_{14}+p_2 q_{15},\notag \\
&p_{15} q_1+p_{13} q_3+p_{10}
q_4+p_8 q_6+p_6 q_8+p_4 q_{10}+p_3 q_{13}+p_1 q_{15},\notag \\
&p_{14} q_1+p_{13}
q_2+p_{11} q_4+p_8 q_7+p_7 q_8+p_4 q_{11}+p_2 q_{13}+p_1 q_{14}\Big\} \,.
\end{align}
The respective probabilities of occurrence are
\begin{align}
	\Pr(\acm(\vec{r}, \vec{v}))= \sum_i \acm_i(\vec{r}, \vec{v}) \,\,, \,\,\,	\Pr(\cm(\vec{r}, \vec{v}))= \sum_i \cm_i(\vec{r}, \vec{v}) \,,
\end{align}
where $\acm_i (\vec{r}, \vec{v})$ and $\cm_i (\vec{r}, \vec{v})$ denote the $i^\textrm{th}$ element of $\acm(\vec{r}, \vec{v})$ and $\cm(\vec{r}, \vec{v})$, respectively. It can be verified that $\Pr(\acm(\vec{r}, \vec{v}))+\Pr(\cm(\vec{r}, \vec{v}))=1$, as expected.

\end{document}